# Wave Excitation and Dynamics in Non-Hermitian Disordered Systems

Yiming Huang,[1,2] Yuhao Kang,[1,2] and Azriel Z. Genack[1,2]

[1] Department of Physics, Queens College of the City University of New York,
Flushing, NY 11367, USA

[2] Department of Physics, The Graduate Center of the City University of New York,
New York, NY 10016, USA

Dynamic and steady state aspects of wave propagation are deeply connected in lossless open systems in which the scattering matrix is unitary. There is then an equivalence among the energy excited within the medium through all channels, the Wigner time delay, which is the sum of dwell times in all channels coupled to the medium, and the density of states. But these equivalences fall away in the presence of material loss or gain. In this paper, we use microwave measurements, numerical simulations, and theoretical analysis to discover the changing relationships among the internal field, transmission, transmission time, dwell time, total excited energy, and the density of states in with loss and gain, and their dependence upon dimensionality and spectral overlap. The field, including the contribution of the still coherent incident wave, is a sum over modal partial fractions with amplitudes that are independent of loss and gain. The total energy excited is equal to the dwell time. When modes are spectrally isolated, the energy is proportional to the sum of products of the modal contribution to the density of states and the ratio of the linewidth without and with absorption. When modes overlap, however, the total energy is further reduced by loss and enhanced by gain. In 1D, the average transmission time is independent of loss, gain, and scattering strength. In higher dimensions, however, the average transmission time falls with loss, scattering strength and channel number, and increases with gain. It is the sum over Lorentzian functions associated with the zeros as well as the poles of the transmission matrix. This endows transmission zeros with robust topological characteristics: in unitary media, zeros occur either singly on the real axis or as conjugate pairs in the complex frequency plane. The transmission zeros are moved down or up in the complex plane by an amount equal to the internal rate of decay or growth of the field. In weakly absorbing media, the spectrum of the transmission time of the lowest transmission eigenchannel is the sum of Lorentzians due to transmission zeros plus a background due to far-off-resonance poles. As the scattering strength increases, conjugate pairs of zeros are brought to the real axis and converted to two single zeros which are constrained to move on the real axis until they combine and leave the real axis as a conjugate pair. The average density of transmission zeros in the complex plane is found from the fall of the average transmission time with absorption as zeros are swept into the lower half of the complex frequency plane. As a sample is deformed, two single zeros and a conjugate pair of zeros may interconvert on the real axis with a square root singularity in the sensitivity of the spacing between transmission zeros to structural change. Thus, the disposition of poles and zeros in the complex frequency plane provides a framework for understanding and controlling wave propagation in non-Hermitian systems.

## I. INTRODUCTION

The scope of studies of resonance phenomena has expanded from the early interest in water waves, musical instruments, tides, pendulums, and catastrophic bridge collapses to encompass the entirety of the physical world described by classical and quantum wave equations. Thus, resonances are central to the study of thermodynamics, electronics, and optics, and determine the interactions of waves and particles across length scales from nuclei, atoms, and molecules to electronic conductors, photonic devices, chaotic cavities, random media, and black holes [1–25].

The modes of a closed system form a complete orthogonal set of undamped oscillatory solutions of the wave equation with real eigenvalues giving point spectra . Once a wave in a medium is coupled to its surroundings through the boundaries, however, the modes of the closed system are broadened into quasinormal modes (QNMs). When it will not lead to confusion, we will refer to resonances simply as modes. The coupling between external channels and modes in lossless systems has been incorporated in random matrix theory in the Feshbach [2] and Heidelberg [3,7,8] approaches and in coupled mode theory [6]. In this work, we seek to discover the relationships between the scattering coefficients, internal fields, total energy excited, dynamics of transmission and the density of states (DOS) in random systems with loss and gain.

The scattering of waves between the channels coupled to a medium is described by the scattering matrix (SM) [3,7,8]. In studies of wave propagation in random one-dimensional (1D), quasi-1D, and slab geometries, it is natural to distinguish between the channels on the two sides of the sample. 1D media include waves on a line, single-mode waveguides or layered media with parallel interfaces. Quasi-1D samples have reflecting transverse boundaries and constant cross-sectional area A. The number of channels on either side of a large sample is $N \sim A/(\lambda/2)^2$ multiplied by the multiplicity due to spin or polarization. The SM then separates into four sectors, $S = \begin{pmatrix} t & r' \\ r & t' \end{pmatrix}$, where $t(t')$ and $r(r')$ are the transmission and reflection matrices for a wave entering from the left (right). In unitary reciprocal, $N$-channel systems, the SM can be retrieved from the transmission matrix (TM).

There has been long-standing interest in Fano resonances in which the scattering spectrum drops sharply and asymmetrically to zero as a result of interference between a narrow resonance and a broad mode or continuum [5]. The Fano resonance was first understood in the study of inelastic electron scattering but is observed in diverse contexts. Recently there has been great interest in applications of Fano resonances to sensing, filtering, and switching in photonic metamaterials [5,14,26,27]. It has been shown recently that reflection can be completely suppressed in all [28–33] or some [34,35] channels of the SM. In coherent perfect absorption (CPA), a wave is wholly absorbed when a zero of the SM is brought to the real axis of the complex energy plane by adjusting the loss and/or the internal structure of the system [28–32,36]. The imaginary axis of the complex plane is the field decay rate. The zeros and poles of the SM are conjugate pairs in conservative systems with the zero in the upper half and the pole in the lower half of the complex energy or frequency plane [28,34]. The poles and zeros move down and up when loss or gain are introduced. CPA is the time-reversal of lasing [28–30]. Zero reflection can be achieved in any set of input channels by bringing a zeros of the reflection matrix (RM) to the real axis by modifying the losses or internal structure of a system [34]. In this work, we consider poles and zeros of the TM [37] in media in which the SM is or is not unitary, which we refer to as unitary or nonunitary media.

The QNMs are solutions of the wave equation with outgoing boundary conditions with complex eigenvalues or poles in the lower half of the complex frequency plane, $\lambda_m = \omega_m - i\gamma_m$. Here $\omega_m$ and $\gamma_m$ are the central frequency and decay rate of the field of the $m^{th}$ mode. $\gamma_m$ is half the decay rate of modal energy and equals the half-width of the Lorentzian spectral line, $\gamma_m = \Gamma_m/2$. If we assume that the modes are not changed by absorption, the field decay rate in the $m^{th}$ mode in a sample with uniform field absorption rate $\gamma$ is $\gamma_m = \gamma_m^0 + \gamma$, where $\gamma_m^0$ is the field decay rate in the lossless medium due to leakage through the boundaries. In an amplifying medium, $\gamma$ is negative. Below the lasing threshold of $\gamma = -\gamma_m^0$, the modal linewidth is $\gamma_m = \gamma_m^0 + \gamma = \gamma_m^0 - |\gamma|$. The contribution of a mode to the DOS at a specific frequency is equal to the value of the modal Lorentzian function at that frequency.

Many key results for closed Hermitian systems have been demonstrated theoretically to carry over to finite lossless systems open at one end [11]. When there is a step in the potential or dielectric constant demarcating the sample from the surroundings in which the potential or dielectric constant is uniform so that waves do not scatter back into the sample, modes form a complete biorthogonal set. Departures from orthogonality grow as spectrally overlap of modes increases and neighboring modes become correlated [9,10,38]. However, the field within the medium can still be expressed as a superposition of modes [9,14]. Because the ratio of the wavelength to the sizes of elements of the internal structure and the boundaries changes with frequency, the coupling to modes of the sample and channels may change with frequency. This gives rise to non-resonant contributions to the SM. [39] But for high-Q resonances in random media with linewidth much smaller than the driving frequency, the coupling to modes is largely independent of frequency.

The relationships between emission and excitation and between the time and frequency domains lie at the heart of the study of wave propagation. The spontaneous emission from a point within a medium to all freely propagating states is proportional to the local DOS (LDOS) [40]. In unitary systems, the inverse of the process of spontaneous emission is the delivery of energy to a point. The sum of the energy delivered to a point for unit flux over all channels, $U(r,\omega)$, is proportional to the LDOS, $U(r,\omega) = 2\pi\rho(r,\omega)$ [41]. The LDOS is also proportional to the imaginary part of the Green's function for return to the point, $\rho(r,\omega) = -\text{Im}G(r,\omega)/\pi$ [12]. The integral of $U(r,\omega)$ over the volume of the sample gives the DOS, $U(\omega) = \int_V U(r,\omega)dr = 2\pi\rho(\omega)$.

The time to scatter between channels is given by the derivative of the phase of the transmitted wavefunction, $\hbar\frac{d\varphi}{dE}$, or classical field, $\frac{d\varphi}{d\omega}$, [42–49,8,50–55]. The scattering time is the temporal average of the delay of the scattered wave packet or pulse between channels in the limit of vanishing pulse bandwidth and diverging pulse length [51]. In unitary media, the Wigner time delay, $\tau_W$, for a system with $M$ channels, is the sum over the $M$ eigenvalues, $\tau_i$, of the Wigner-Smith delay matrix, $Q = -iS^\dagger\frac{dS}{d\omega}$ [43]. This sum is proportional to the DOS, which is a sum over Lorentzian modal contributions [8]. It is also equal to the dwell time $\tau_D$, which is the sum of the travel time of the particle or wave between all pairs of channels weighted by the corresponding flux transmission coefficients. The Wigner time is also proportional to the number of particles [47,56], which corresponds to the energy of classical waves excited in the medium.

$$\tau_W \equiv \sum_1^M \tau_i = \tau_D = 2\pi\rho = 2\sum_1^M \frac{\gamma_m^0}{(\omega-\omega_m^0)^2+(\gamma_m^0)^2} = U, (\gamma = 0). \quad (1)$$

In quasi-1D samples, $M = 2N$.

In unitary media, $\rho$ and $U$ may also be expressed in terms of the transmission time, $\tau_T$, which is the sum of transmission times over the eigenchannels of the TM. The TM can be expressed via the singular value decomposition as $t = U\Lambda V^\dagger$, where $V$ and $U$ are unitary matrices and $\Lambda$ is a diagonal matrix. The columns of $V$ and $U$, $\boldsymbol{v}_n$ and $\boldsymbol{u}_n$, are the transmission eigenfunction for the $n^{th}$ transmission eigenchannel at the incident and outgoing surfaces of the sample. The squares of the diagonal elements of $\Lambda$ are the transmission eigenvalues, $\tau_n$. These are the eigenvalues of $tt^\dagger$, which are the flux transmission coefficients of the transmission eigenchannels [57–64]. The transmission eigenvalues are indexed from 1 to $N$ in order of decreasing transmission. The transmittance is the sum of transmission eigenvalues, $T = \sum_{n=1}^N \tau_n$ [57–59]. The transmission time for the $n^{th}$ transmission eigenchannel is the spectral derivative of the composite phase shift accumulated in transmission, $t_n = \frac{d\theta_n}{d\omega} = \frac{1}{i}\left(\boldsymbol{u}_n^* \cdot \frac{d\boldsymbol{u}_n}{d\omega} - \boldsymbol{v}_n^* \cdot \frac{d\boldsymbol{v}_n}{d\omega}\right)$ [54]. Here $\boldsymbol{v}_n$ and $\boldsymbol{u}_n$ are the $n^{th}$ columns of $V$ and $U$, which are expressed in terms of the incident and outgoing channels, respectively, and $\theta_n$ is the difference between the phase averaged over the flux in the channels contributing to the $n^{th}$ eigenchannel on the output and input surfaces of the sample. The transmission time in a unitary medium is then the sum of eigenchannel transmission times

$$\tau_T \equiv \sum_{n=1}^N t_n = \pi\rho = \sum_m \frac{\gamma_m^0}{(\omega-\omega_m)^2+(\gamma_m^0)^2} = U/2, \gamma = 0. \quad (2)$$

The wave within a unitary sample can be decomposed into a sum of modal contributions each of which is a product of a spatially varying amplitude, $a_m(\boldsymbol{r})$ and a resonance partial fraction. The resonance partial fraction is proportional to the Fourier transform of the modal field arising from delta function excitation, which is an exponential decaying oscillation, $E_m(t) \sim \cos\omega_m t \exp(-\gamma_m t)$. The field inside a 1D sample and the transmission are thus given by [11]

$$E(x) = \sum_m \frac{a_m^0(x)}{\omega-(\omega_m^0-i\gamma_m^0)}, \quad (3a)$$

where $a_m^0(x)$ is the amplitude of the mode at a depth $x$ into the unitary medium, and the transmission coefficient in a sample of length $L$ is

$$T = |t|^2 = \left|\sum_m \frac{a_m^0(L)}{\omega-(\omega_m^0-i\gamma_m^0)}\right|^2. \quad (3b)$$

Equivalent relations hold for the TM and for the transmittance in multichannel systems.

The DOS relates dynamic and static aspects of transport in unitary media. The LDOS and hence rate of spontaneous emission may be enhanced or suppressed relative to free space in cavities [65,66], photonic crystals [67–69], metamaterials [70], and near surfaces [71]. The vanishing of the DOS in electronic band gaps and the implementation of impurity bands within the gap makes possible the control of charge density and transport in semiconductors [72]. The exclusion of light from photonic band gaps of photonic crystals due to the suppression of the DOS [67] is accompanied by the suppression of spontaneous emission [67]. At the same time, the enhancement of the DOS at the band edge of photonic crystals gives rise to a lengthened modal dwell time suppresses the laser threshold [73]. Light harvesting can be enhanced by broadband, wide-angle absorption in high-Q slow modes in 2D photonic crystals [74]. The DOS may also be enhanced in hyperbolic metamaterials in which ultralow mode volumes at large wavevectors enable low volume photonic devices [70]. The exclusion of light from the band gap of 2D periodic photonic materials enables efficient transmission in photonic crystal waveguides with sharp bends [75]. Similarly light in a photonic topological insulator is excluded from the band gap in the bulk of the material but propagates unimpeded along the edge in a chiral edge state [76]. The DOS also provides a basis for describing transport in biological realms [77,78]. The average time spent by insects within a given region as they forage for food by executing a random walk parallels the average time of diffusing waves within a unitary medium. This corresponds to $\tau_W$, which is proportional to the DOS, and so is independent of the mean free path [77].

The understanding of the scaling of the statistics of wave propagation in disordered systems is built upon the linkage between static and dynamic aspects of transport and the DOS. The statistics of transport in nondissipative open random media are determined by the Thouless number. The Thouless number is equivalent to the degree of modal overlap, which can be expressed as $\delta = \delta\omega/\Delta\omega$, where $\delta\omega$ is the average linewidth and $\Delta\omega$ is the average spacing between the modes of the open medium [79]. The Thouless number may also be expressed as the ratio of the Heisenberg and Thouless times, $\delta = t_H/t_{Th}$. Here $t_H = 1/\Delta\omega$, which is the average of the DOS and equals the time required to visit each coherence volume of the sample, and $t_{Th}$ is the typical time for the energy in a mode to leak through the sample boundaries. In unitary systems, the Thouless number is equal to the dimensionless conductance, $g$, which is the conductance in units of $e^2/h$, and equal to the transmittance, $\delta = g = \langle T \rangle$ [79]. This reflects the link between steady-state and dynamic transport. The localization threshold in open systems lies at $\delta = 1$ [79,80]. Modes of the medium are localized [81] when they are spectrally isolated, $\delta < 1$, but extend throughout the medium when they overlap, $\delta > 1$. The spectral narrowing of spatially localized modes due to weak coupling to the boundaries can be seen in spectral lines in transmission [82]. The role of dissipation is crucial in the study of Anderson localization because even weak absorption dramatically affects transport through long-lived localized states. As dissipation increases, $\langle T \rangle$ falls but $\delta$ increases since modes are broadened by loss so that $\langle T \rangle$ and $\delta$ are no longer equal.

In this work, we utilize microwave measurements, numerical simulations and theoretical analysis to explore the power and limitations of the descriptions of static and dynamic aspects of wave propagation in random media with uniform loss or gain in terms of coherent and incoherent sums of modes. We demonstrate in strongly absorbing 1D samples that the field excited by an incident wave from the left (right) with unit flux is the coherent sum of products of partial fractions, $\frac{1}{\omega - \lambda_m}$, and modal amplitudes $a_m(x)$ ($a'_m(x)$), which are the left (right) biorthogonal states. The modal amplitudes, as well as the resonant frequencies are not changed up to the high levels of loss or gain studied here. The spectral range of modes needed to represent the field at a point increase rapidly as the incident surface is approached. We find the way the relationships between $\tau_T$, $\rho$, and $U$ are transformed in samples with loss or gain from the proportionality between these variables in lossless media as expressed in Eq. (2).

In non-Hermitian 1D media and in unitary multichannel systems, the transmission time is proportional to the DOS, which is the incoherent sum of normalized Lorentzian functions associated with poles. The average of $\tau_T$ over a random 1D ensemble is independent of the strengths of scattering, loss, or gain and is given by the ballistic time $t_+ = \frac{L}{v_+}$. In higher dimensions, however, $\tau_T$ depends on scattering strength and loss. It is a sum of Lorentzian functions associated with both the poles and zeros of the determinant of the TM. In unitary systems, transmission zeros can appear only as single zeros on the real axis or as conjugate pairs in the complex frequency plane [37]. When uniform loss or gain are added, the transmission zeros along with the poles move down or up by $\gamma$. As a result, transmission may vanish in systems with and without loss and gain.

In quasi-1D media, $\langle \tau_T \rangle$ falls with increasing absorption as zeros are brought to the lower half of the complex plane. In diffusive multichannel media, the ordering of transmission times of transmission eigenchannels $t_n$ crosses over with increasing loss from longer times being associated with higher transmission eigenvalues in conservative systems to the reverse in in strongly absorbing samples. The probability distribution of the imaginary part of the transmission zeros in the complex plane is found from the drop in $\langle \tau_T \rangle$ with absorption. We present the evolution of zeros in systems in which the disorder is increased systematically. In unitary systems, conjugate pairs of zeros may move towards the real axis and meet at a zero point (ZP) at which the pair is transformed into two single zeros [37]. The sensitivity of the distance between transmission zeros in the complex plane has a square-root singularity at the ZP.

This paper is organized as follows: The introduction presents an overview of waves scattering from and within conservative random media and outlines the paper. Section II presents the DOS in terms of the counting number, which in 1D is related to both the number of half wavelengths across the sample and the spectral derivative of the phase. The simulations presented in this section establishes the modal basis for the changing nature of spectra and probability distributions of $\tau_T$ with scattering strength in lossless 1D systems. Section III begins with a discussion of microwave measurements of the average energy density $\langle U(x) \rangle$ and of the probability distribution of $\tau_T$ in random ensembles of single-mode waveguides with and without loss. Simulations in random layered media over a range of dissipation show that while $\langle U(x) \rangle$, $\langle U \rangle$, the Wigner time, and the reflection time are suppressed by absorption and enhanced by gain, $\langle \tau_T \rangle$ is unchanged by the introduction of dissipation and gain. Theoretical calculations of $E(x)$, $T$, $U$, and $\tau_T$ in random 1D systems with and without loss and gain are presented and compared to simulations of spectra of these quantities in a random layered sample. As the spectral window in increases, all these quantities converge to the appropriate expression in term of modes demonstrating the effective completeness of QNMs. Section IV presents the statistics of the $t_n$ and their sum $\tau_T$ with changing absorption, as well as the statistics of transmission zeros with changing strengths of scattering and absorption. The changing disposition of transmission zeros in the complex plane in a sequence of samples with increasing scattering strength is given. This shows the divergence of the displacement of transmission zeros in the complex plane with change in the sample configuration near a ZP. The sharp spectra of transmission and of transmission time of the lowest transmission eigenchannel near transmission zeros are shown to be, respectively, products of quadratic functions and as a sum of Lorentzian lines. As for in 1D samples, modal expressions for propagation variables converge to the predicted sum in terms of the modes and transmission zero as the frequency range analyzed increases. We conclude in Sec. V with an overview of the broadened perspective of wave propagation in random media achieved in this work and a discussion of open questions and possible applications.

## II. TRANSMISSION TIME AND DENSITY OF STATES IN UNITARY RANDOM 1D SYSTEMS

The statistics of transmission time in unitary random systems is connected to the DOS. The DOS of a volume with uniform index of refraction is the spectral derivative of the counting number, $\rho(\omega) = dN(\omega)/d\omega$, where $N(\omega)$ is the number of modes with angular frequency below $\omega$. In a uniform medium enclosed in a volume $V$ in $d$ dimensions, the number of states with a specific spin or polarization with frequency below $\omega$ and wavelength longer than $\lambda = 2\pi v/\omega$ is $N(\omega) \sim V/(\lambda/2)^d$, in line with Weyl's law. In an inhomogeneous closed medium, the average of the DOS should be related to the effective wavelength, which can be obtained from the correlation length of the field. The separation at which the field correlation function first crosses zero is $\lambda/2$ [83,84]. This defines the effective wavelength in the medium, in terms of which $N(\omega) = V/(\lambda/2)^d$. In principle, the DOS, $\rho(\omega) = \frac{dN}{d\omega} = \frac{N(\lambda)}{d\lambda}\frac{d\lambda}{d\omega} = \frac{V}{2(\lambda/2)^2}\frac{d\lambda}{d\omega}$, varies between systems because of differences in wave dispersion. However, in random systems without strong structural correlation, the DOS is the same as in a uniform system with the same effective wavelength. Since the ensemble average of the DOS in a random unitary system depends only on the wavelength and the sample dimensions, the average of the transmission time, $\langle \tau_T \rangle = \pi \langle \rho \rangle$ would be expected to be independent of the scattering strength. Since the wavelength is determined by the real part of the dielectric constant, the average DOS should also be independent of absorption. The average transmission time through a random medium should therefore be the same as the transmission time in the equivalent uniform lossless medium. Thus, in 1D, $\langle \tau_T \rangle = \frac{L}{v_+} \equiv t_+$, and in quasi-1D

$$\langle \tau_T \rangle = N\frac{L}{v_+} \equiv Nt_+ \quad (4)$$

where, $\frac{1}{N}\sum_{n=1}^{N}\frac{L}{v_n} \equiv \frac{L}{v_+}$, and $v_n$ is the longitudinal velocity associated with the $n^{\text{th}}$ propagating transmission eigenchannel coupled to the medium. Thus $\tau_T$ is the time to visit each coherence time in the medium, which is the Heisenberg time.

The independence of $\langle \tau_T \rangle$ upon scattering strength in random 1D unitary systems is demonstrated in the transfer-matrix simulations shown in Fig. 1. A plane wave of wavelength 3 mm is normally incident on a binary dielectric stack shown schematically in Fig. 1(a). The sample is composed of $N_{\text{layer}} = 200$ layers with total length $L = 1$ m and with layer thickness selected randomly from a uniform distribution [0,10] mm. The real indices of refraction, $n_1$ and $n_2$, alternate between layers with indices of refraction $1 \pm \Delta n$. The sample is surrounded by free space with unit index of refraction. The values of the indices of refraction in the sample for a desired value of $L/\ell$ in the layered sample is shown in Appendix A to be set by the relation, $N_{\text{layer}} \ln[n_1(2 - n_1)] = -L/\ell$, up to a small correction due to the weaker scattering at first and last interfaces of the sample. The value of $L/\ell$ given by this relation for each ensemble is within 0.3% of the value obtained in the linear fit to the average of the logarithm of the energy density in the sample when the sample is excited by unit flux from the left side, with $<\ln u(x)v_+> = -x/\ell$ [85], shown in Fig. 1(b).

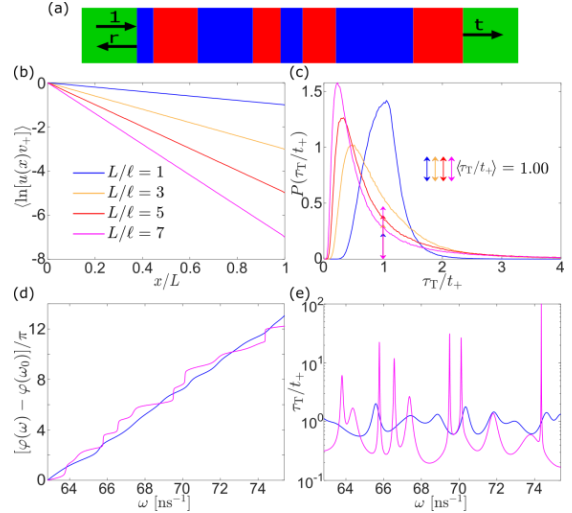

FIG. 1. Transmission time and phase in lossless random media. (a) Schematic diagram of the random layered 1D medium. The alternating blue and red dielectric layers have refractive indices of refraction of $1 \pm \Delta n$, while the index of the green surrounding medium is 1. (b) The linear falloff of the logarithm of the average energy density for unit flux incident from the left in samples with different scattering strengths gives the mean free path, $\langle \ln u(x)v_+ \rangle = -x/\ell$. (c) The probability distribution of transmission time for the same ensembles as in (b). The up-down arrow indicates the average transmission time, which is independent of scattering strength, $\langle \tau_T \rangle = t_+$. (d) Spectra of phase of the transmitted field for samples with $L/\ell$ of 1 and 7. The total increase in phase in this frequency range for these two configurations changes little with scattering strength. $\omega_0$ is the smallest value of the $\omega$ in the window. (e) Spectra of the transmission time, $\tau_T = d\varphi/d\omega$, corresponding to the spectra of the phase in (d). Spectral features become sharper as the scattering strength increases.

The probability distributions of $\tau_T/t_+$ for the same ensembles for which the values of the localization parameter $s = L/\ell$ are found in Fig. 1(b) are shown in Fig. 1(c). The distribution can be seen to broaden with increasing scattering strength, however, the average transmission time is the same for all scattering strengths, including uniform samples without scattering, so that $\langle \tau_T \rangle = L/v_+ = t_+$. The reason for the higher probability of long and short transmission times in samples with higher scattering strength can be seen in the typical spectra of the phase and phase derivative shown in Figs. 1(d) and 1(e) for two random configurations with the smallest and largest scattering strengths. In sample with weak scattering, for which $\delta > 1$, modes are wide, and modal overlap is strong. $\varphi(\omega)$ is, therefore, a smooth function with small departures from a line with slope $\langle d\varphi(\omega)/d\omega \rangle = t_+$. In contrast, in a more strongly scattering sample, in which modal overlap is weak, $\delta < 1$. The resonances are then narrow so that the phase increases rapidly near resonance where the DOS and $\tau_T$ are large and slowly between resonance. As the resonances narrow, higher values of $\tau_T/t_+$ are reached on resonance so that the tail of $P(\tau_T/t_+)$ extends to higher values of $\tau_T/t_+$ while the bulk of the distribution of $\tau_T/t_+$ shifts to small values found between resonances.

## III. PROPAGATION IN NON-UNITARY RANDOM 1D SYSTEMS

### A. Measurement of energy density and transmission time in absorbing single-mode random waveguides

Measurements of field spectra along the length of random layered samples contained in a waveguide supporting a single transverse mode are carried out in an ensemble of 100 random sample configurations with use of a vector network analyzer. Measurements are made in the frequency range $10.00 - 10.70$ GHz in a copper waveguide with cutoff frequency of 6.56 GHz. The medium inserted in the waveguide is composed of randomly positioned ceramic elements with index of refraction $n = 1.7$ and U-channel Teflon spacers. Photos of a portion of the waveguide and of the scattering medium inserted into the waveguide are shown in Figs. 2(a) and 2(b), respectively. Measurements of the intensity inside the sample shown in Fig. 2(c) are carried out in samples of length $L = 86$ cm, while measurements of the spectral derivative of the field $\tau_T = d\varphi/d\omega$ shown in Fig. 2(d) are made in a shorter sample of length $L = 80$ cm. The shorter sample is used to provide additional space for measurements of the field before the sample to more accurately determine the incident field and so the transmission time. The wave is detected by an antenna inserted sequentially into small holes spaced by 1 cm along the waveguide. The measured field is normalized to the incident field, which is found by fitting the expression for the sinusoidal variation of the intensity due to counter-propagating incident and reflected waves at 5 points in front of the sample to the measured intensity. The sample is weakly absorbing with an absorption rate of $9.4 \times 10^{-3}$ ns$^{-1}$ corresponding to $\gamma = 4.7 \times 10^{-3}$ ns$^{-1}$ determined from the linewidth of modes localized in the middle of the sample with reflectors placed at the ends. This corresponds to an absorption length of $\ell_a = v_+\tau_a = 2v_+/\gamma = 23$ m, giving $L/\ell_a = 0.035$. The impact of the weak absorption in this sample is removed by Fourier transforming field spectra into the time domain, compensating for the average decay due to absorption by multiplying by a factor $e^{\gamma t}$, and then Fourier transforming back into the frequency domain, as discussed in Appendix B. Following this approach, additional absorption is effectively incorporated into the sample by multiplying the field in the time domain by a decaying exponential factor. The energy densities within the sample and the probability distributions of transmission times for the random ensemble with the impact of the small natural absorption removed and with added dissipation of $\gamma = 0.11$ ns$^{-1}$ are shown in Figs. 2(c) and 2(d), respectively.

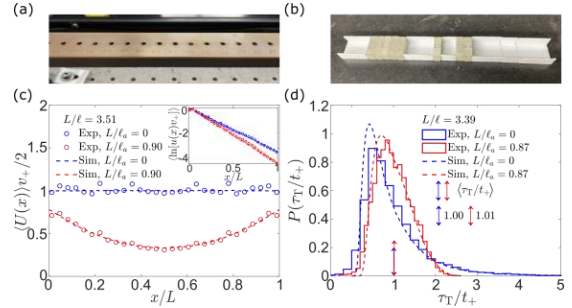

FIG. 2. Microwave measurement of energy density and transmission time. (a) Photo of the single-mode copper waveguide with holes separated by 1 cm. The field is measured by an antenna inserted in the holes. (b) Photo of one segment of the sample showing dielectric slabs and U-channel spacers. (c) The average energy density in 100 random configurations in a sample of length $L = 86$ cm in a single-mode random waveguide. The blue/red circles give the average energy density in samples without ($L/\ell_a = 0$) and with absorption ($L/\ell_a = 0.90$). The dashed lines are simulations for the same values of $L/\ell$ and $L/\ell_a$ as in the experiment. Measurements of $\langle \ln u(x) v_+ \rangle$ from which $\ell$ and $\ell_a$ are found are shown in the inset. (d) The probability distribution function for the transmission time for the samples with the same local disorder as in (c) with length, $L = 80$ cm. The solid/dashed curve shows the transmission time for the experiment/simulation. The up down arrows indicates the average values of $\tau_T/t_+$ ($t_+ = 3.70$ ns).

The average energy density within the sample for excitation with unit flux from the left, is $\langle u(x)\rangle = \frac{1}{2}\langle\varepsilon\rangle E^2$. The average dielectric constant, $\langle\varepsilon\rangle = 1.2$, is used because the precise position of the dielectric slabs in the sample is not known. Since the disorder in the random structure is statistically homogeneous throughout the sample, the average energy density in the sample when excited from the left and right, $\langle u(x)\rangle$ and $\langle u'(x)\rangle$, respectively, will be symmetrical with respect to the center of the sample, $\langle u'(x)\rangle = \langle u(L-x)\rangle$. Adding the average energy density for excitation from the left and right gives $\langle U(x)\rangle = \langle u(x)\rangle + \langle u(L-x)\rangle$. This is shown in Fig. 2(a) as blue/red circles for the lossless/absorbing sample. $\langle U(x)\rangle$ in the sample without loss is uniform, as expected, since the average of the LDOS is uniform. In the absorbing sample, $\langle U(x)\rangle$ falls towards the middle of the sample since waves reaching the center of the sample have spent more time within the sample on average than the waves near the sample boundaries and are consequently more strongly absorbed.

We determine the mean free path, $\ell$, and the absorption length, $\ell_a$, from the relation $\langle \ln u(x) v_+\rangle = -x/\ell - x/\ell_a$ using a linear fit to the data shown in the inset of Fig. 2(c). This gives, $\ell = 24.5$ cm and $\ell_a = 107$ cm, and the dimensionless parameters $s = L/\ell = 3.51$, $L/\ell_a = 0.81$. The values of the localization parameter in the sample with $L = 80$ cm for which results are shown in Fig. 2(d) is $s = 3.39$. Simulations carried out with these parameters give the thin curves in Fig. 2(c), which closely match the measurements of $\langle U(x)\rangle$. Absorption is incorporated in the simulations by adding an imaginary component $n_i$ to the indices of refraction $n_1$ and $n_2$ such that $2\frac{\omega}{c}n_i = 1/\ell_a$.

The probability distributions of transmission time measured in the random ensembles with and without absorption are shown in Fig. 2(d). In the sample in which the impact of absorption is removed, we find $\langle\tau_T\rangle = 3.70$ ns. The probability distribution at high and low values is greater in the sample without absorption. This is consistent with the increase in the probability distribution of the transmission time delay at long and short times in more strongly scattering samples with narrowed average linewidths shown in Fig. 1(c). The average transmission time in the two ensembles with different modal broadening are the same within the uncertainty in the measurements of 1%. A small fraction of the measured transmission times is negative. Negative transmission times in 1D do not occur in the simulations, but they occur arise when the measured transmitted field is comparable to the noise level. The difference in $t_+$ between the sample without absorption and with the same absorption as in the experiment is less than 0.2%. The results shown in Figs. 1 and 2 indicates that the average transmission time through a random ensemble of 1D sample is independent of the strengths of scattering and absorption so $\langle\tau_T\rangle = t_+ = L/v_+$.

## B. Simulations of total energy and delay times in nonunitary random 1D media

To better understand the nature of propagation in random media, we carry out simulations of total energy excited within the sample and of key delay times over a wide range of absorption and gain. In the simulations shown in Figs. 3(a) and 3(b) carried out in ensembles of random 1D samples with the same value of $s$ of 3.39 as in the measurements in Fig. 2, the averages of the energy density $\langle U(x)\rangle$ and of the total excitation within the sample $\langle U\rangle$ are suppressed by loss and enhanced by gain. The total energy within the sample can be found by equating the rate of loss of energy due to absorption within the sample, $2\gamma U$, to the difference between the sum of flux into the sample on both side of 2 for unit flux, and the flux out of the sample to give [86]

$$U = U_1 = [2 - (|t|^2 + |t'|^2 + |r|^2 + |r'|^2)]/2\gamma. \quad (5)$$

$U$ can also be found by integrating the energy density in a 1D medium excited by unit incident flux from the left and right. The energy density is $U(x) = u(x) + u'(x)$, with $u(x) = \frac{1}{2}\varepsilon(x)E(x)^2$ and $u'(x) = \frac{1}{2}\varepsilon(x)E'(x)^2$ in

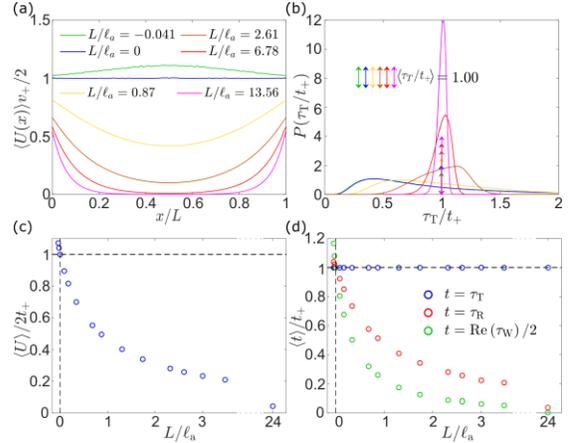

FIG. 3. Simulations of energy density and propagation times in random 1D media with $L/\ell = 3.39$. (a) The average energy density for different values of loss and gain in samples with the indicated values of $L/\ell_a$. (b) The probability distribution of transmission time for the values of $L/\ell_a$ indicated in (a). The up-down arrows indicate the values of the normalized average transmission time, which are independent of loss or gain. (c) The average energy excited within the sample vs. strength of absorption or gain. (d) The ensemble average of the transmission time (blue circles), reflection time (red circles) and the real part of the Wigner time (green circles) for different strengths of absorption or gain. $\langle\tau_T\rangle$ is invariant up to the highest level of absorption $\gamma = 3.6$ ns$^{-1}$ at which $L/\ell_a = 24$.

terms of the field due to excitation from the left and right, respectively. The total excited energy is
$U = U_2 = \int_0^L u(x) + u'(x) dx = \int_0^L \frac{1}{2}\varepsilon(x)[E(x)^2 + E'(x)^2]dx.$ (6)

Whereas, $\langle U \rangle$ is independent of scattering strength for a given value of $t_+$ in unitary systems but falls and rises with increasing absorption and gain, $\langle \tau_T \rangle$ is pegged to $t_+$ up to the largest values of $L/\ell_a = 24$ at which it was feasible to determine the phase derivative of the transmitted field, as seen in Figs. 3(c) and 3(d). At this level of absorption, $\langle \ln T \rangle = -L/\ell - L/\ell_a = -3.39 - 24 = -27.4$. The average of the field decay rate due to absorption in the layered sample with $v_+ = c$ is then $\gamma = \frac{1}{2}\Gamma_a = \frac{1}{2}\frac{c}{\ell_a} = 3.6$ ns$^{-1}$. Up to this high level of absorption, the average transmission time is proportional to the average density of states, $\langle \tau_T \rangle = t_+ = \pi \langle \rho \rangle$. The probability distribution of $\tau_T$ relative to the transit time in the uniform system, $t_+$, $P(\tau_T/t_+)$, narrows as $L/\ell_a$ increases (Fig. 3(c)). This is due to the increasing width of the modes of the medium, which leads to a smoother variation of $\varphi$ and a narrower range of values of the spectral derivative.

The Wigner time delay becomes complex in nonunitary systems and is therefore not well-defined. Two scattering times that can be explored are the real part of the Wigner time and the reflection time, as functions of absorption. Both these times are seen in Fig. 3(d) to rise in amplifying media and fall in absorbing media. The invariance of $\langle \tau_T \rangle$ is consistent with $\tau_T$ being a sum over modes and the constant value of the average DOS for fixed $t_+$. The contribution to $\langle \tau_T \rangle$ of a single mode within a spectral range $\overline{\Delta \omega}$ much greater than the mode linewidth is the ratio of the change in phase of the field due to the mode of $\pi$, which is independent of loss or gain, and the spectral range, giving $\pi/\overline{\Delta \omega}$. Since the number of modes in the medium is not changed by the introduction of loss or gain, the modal contribution to $\langle \tau_T \rangle$ is constant.

In lossless systems, the Wigner time is equal to the dwell time, $\tau_D$, which is the sum over all channels of the residence time of waves within a medium, and to $U$, $\tau_W = \tau_D = U$. In the presence of loss, $\tau_W$ becomes complex and its real part is not equal to $U$, as seen in Fig. 3, but $\tau_D$ is still well defined. The contribution of each channel to $\tau_D$, is the residence time of the wave within the medium before the wave either exits the medium through the boundaries or is absorbed. In 1D,
$\tau_D = T\tau_T + R\tau_R + T'\tau_T' + R'\tau_R' +$
$2\gamma \int_0^L u(\omega,x) \frac{\partial \varphi(E(\omega,x))}{\partial \omega} + u'(\omega,x) \frac{\partial \varphi(E'(\omega,x))}{\partial \omega} dx,$ (7)
where $2\gamma u(\omega, x)$ is the energy dissipation rate at position $x$ and $\frac{\partial \varphi(E(\omega,x))}{\partial \omega}$ is the travel time from the source to $x$ [49]. Simulations of spectra of $U_1$, $U_2$ and $\tau_D$ are shown in Fig. 4 for an 1D sample drawn from an ensemble of 1D systems with $s = L/\ell = 3.39$ with and without loss and gain. In all cases, spectra of $U_1$, $U_2$ and $\tau_D$ overlap, so that $U = U_1 = U_2 = \tau_D$. The equality of $\tau_D$ and $U$ is natural from the particle perspective. For electromagnetic waves, for example, $U$ is proportional to the number of photons in the medium which is proportional to the dwell time of injected photons within the medium, regardless of the strength of scattering and whether photons leave the medium by escaping through the boundaries or by being absorbed. In a uniform, lossless medium, $\tau_D = 2Nt_+ = U$, so that, in general, $\tau_D = U$.

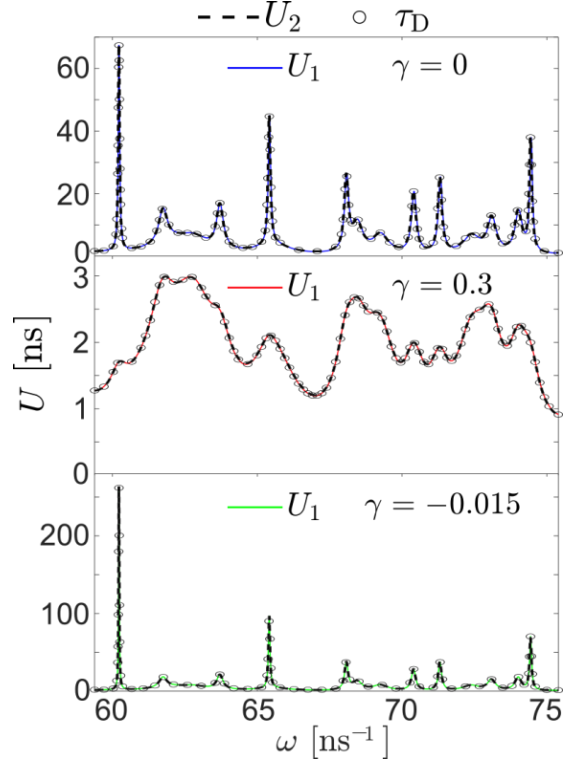

FIG. 4. Spectra of total internal energy and dwell time. Simulations of total internal energy and dwell time without loss or gain ($\gamma = 0$ ns$^{-1}$), with loss ($\gamma = 0.3$ ns$^{-1}$), with gain ($\gamma = -0.015$ ns$^{-1}$) for a single sample drawn from an ensemble with $s = 3.39$. The solid lines are $U_1$, the black dashed lines are $U_2$ and the circles are $\tau_D$.

The independence of the dwell time of a wave within a lossless diffusive medium of scattering strength arises from the relation $\tau_D = U = 2\pi\rho$. But in non-Hermitian systems, $U$ is not proportional to $\rho$. This raises the question of whether $\tau_D$, which is equal to $U$ and proportional to the sum over pathlengths for all channels, is independent of the strength of scattering in samples with the same absorption. The variation of $U$ with scattering strength and loss in 1D and quasi-1D media is explored in Appendix D. For a given rate of dissipation, and so of $\tau_D$, $U$ is seen in Fig. 17 to decrease with scattering strength. Therefore, the average pathlength is not independent of scattering strength in the presence of loss. This conclusion may be applied to the pathlength of randomly walking insects expiring at a fixed rate. Their residence time within a given region will decrease with scattering strength.

### C. Theory of field, energy density and transmission time in nonunitary random media: Poles and zeros

The completeness of the QNMs has been demonstrated theoretically for a 1D dissipationless sample closed at one end and open at the other when there is a step in the wave velocity at the open end and when the potential or dielectric constant is uniform outside the sample so that the wave is not scattered back into the sample [11]. Here, we consider the completeness of modes in 1D and quasi-1D samples open at both ends with uniform loss or gain. The statistics of $\tau_T$ and high-quality modal fits of key variables of propagation are demonstrated in 1D in Sec. IIID and in quasi-1D in Sect. IV.

According to Mittag-Leffler's theorem, the Green's tensor can be expanded as $G(\omega) = \sum_{m=1}^{\infty} \text{Res}[G(\omega_m)](\omega - \omega_m)^{-1}$, where the $\text{Res}[G(\omega_m)]$ is the residue of the Green's function around $\omega_m$ [87]. The spatial variation of $\text{Res}[G(\omega_m)]$ for a source on the left or right side of the sample corresponds to the amplitudes of the biorthogonal states. The Green's function in an open system is given by $G(r', r, \omega) = \langle r' | \frac{1}{E - H_{eff}} | r \rangle = \sum_n \frac{\langle r' | \psi_{mr} \rangle \langle \psi_{ml} | r \rangle}{E - E_m + i\Gamma_m/2}$. Here $\langle \psi_{ml} |$ and $|\psi_{mr}\rangle$ represent the biorthogonal states of the effective Hamiltonian of the scattering region, $H_{eff}$, for excitation from the left or right of the sample. Assuming that the biorthogonal states are not changed by loss or gain, the field within and the flux transmitted through a non-Hermitian sample can be expressed as

$$E(x) = \sum_m \frac{a_m^0(x)}{\omega - (\omega_m^0 - i(\gamma_m^0 + \gamma))}, \quad (8a)$$

$$T = |t|^2 = |\sum_m \frac{a_m^0(L)}{\omega - (\omega_m^0 - i(\gamma_m^0 + \gamma))}|^2, \quad (8b)$$

where $a_m^0(x) = \text{Res}[G(\omega_m, x)] = \langle r' | \psi_{mr} \rangle \langle \psi_{ml} | r \rangle$ is proportional to the modal amplitude for excitation from the left in both unitary and nonunitary media.

#### 1. Total energy

The energy excited in lossless media with unit flux incident in all channels is calculated in Appendix E using the Heidelberg model following the Feshbach quantum formalism [2,3,7,8]. The Heidelberg model treats the coupling of the continuum states of the channels to the modes of the scattering region. The quantum calculation of Appendix E only holds when the Hamiltonian is Hermitian. The result of the quantum calculation may be transcribed for classical waves in a system without loss as

$$U = 2\sum_m \frac{\gamma_m^0}{(\omega - \omega_m)^2 + (\gamma_m^0)^2}, \gamma = 0. \quad (9a)$$

This result is in accord with the relationship between $U$ and $\tau_W$ in Eq. (1) in a unitary system but breaks down in the presence of dissipation. when

The total energy excited in a single isolated mode at $\omega_m$ with linewidth $\gamma_m$ can be calculated by considering the response to $\delta$-function excitation at $t = 0$. The response at a point $x$ in the medium is a damped oscillation, $E_m(x, t) \sim \exp(-\gamma_m t) \cos\omega_m t$, which may be Fourier transformed to give the response in the frequency domain. For a high-Q resonance, $\omega_m \gg \gamma_m$, this gives, $E_m(x, \omega) \sim \frac{1}{\omega - (\omega_m - i\gamma_m)}$. The spectral variation of the energy density would then be $U_m(x, \omega) \sim \frac{1}{(\omega - \omega_m)^2 + \gamma_m^2}$. Since all points in the medium have the same spectral variation, the total energy excited in a spectrally isolated mode is $U_m(\omega) \sim \frac{1}{(\omega - \omega_m)^2 + \gamma_m^2}$. In a lossless medium, the energy excited summed over all incident channels is $\frac{2\gamma_m^0}{(\omega - \omega_m)^2 + (\gamma_m^0)^2}$, in accord with Eqs. (1) and (9a). The integral of $U_m(\omega)$ over frequency is $2\pi$. The integral over frequency of the energy excited in an isolated high-Q mode by $\delta$-function excitation in a system with loss is reduced relative to that in the lossless system by a factor equal to the ratio of the temporal integral of the decaying intensity in the two systems of $\frac{\gamma_m}{\gamma_m^0}$. Since the function with the same spectral variation as the mode in an absorbing medium and with the same integral over frequency would be $\frac{2\gamma_m}{(\omega - \omega_m)^2 + \gamma_m^2}$, the total energy excited in the mode in the lossy system is $U_m = \frac{2\gamma_m}{(\omega - \omega_m)^2 + \gamma_m^2} \frac{\gamma_m^0}{\gamma_m}$. This gives,

$$U_m = \frac{2\gamma_m^0}{(\omega - \omega_m)^2 + \gamma_m^2}, \gamma_m = \gamma_m^0 + \gamma. \quad (9b)$$

In Sec. IIID, we demonstrate the validity of Eq. (9b) in simulations of total energy excited near resonance with a defect state in the band gap of an absorbing photonic crystal which does not overlap strongly with modes of the pass band. Modal overlap is also low in random media in which waves are strongly localized so that $\delta \ll 1$, and when samples are weakly coupled to their surroundings. But generally, the total energy excited in a nonunitary sample is not given by the sum over modes when modes overlap. Nonetheless, $U = \sum_m \frac{2\gamma_m^0}{(\omega - \omega_m)^2 + \gamma_m^2}$ is a good approximation in samples in which modal overlap is weak and a serviceable approximation in weakly absorbing or amplifying media with appreciable modal overlap, as shown in Sec. IIID. Simulations presented in Sec. IIID show that the total excited energy in nonunitary media may be expressed, in terms of the integral over the energy density in Eq. (6) with $E$ expressed as the coherent sum over modes in Eq. (8a).

*2. Transmission time*

To find the transmission time for classical waves in a quasi-1D sample, $\tau_T = \frac{d}{d\omega}\arg\det(t)$, it is necessary to find the determinant of the transmission sector of the SM. Using the Heidelberg model, which gives the coupling between a scattering region and its surroundings [7,8,88,89], the determinant of the TM is found to be [37]

$$\det(t) \sim \frac{\prod_i(\omega-\eta_i)}{\prod_m(\omega-\lambda_m)} \quad (10)$$

The summation is over all QNMs with poles $\lambda_m = \omega_m - i\gamma_m$ and all transmission zeros with zeros at $\eta_i = Z_i + i\zeta_i$ in the complex frequency plane. The phase of $\det(t)$ is the difference between the sum of phases of the factors in the numerator and the denominator of Eq. (10). The transmission time is given by the spectral derivative of the phase of $\det(t)$ gives [37]

$$\tau_T = \tau_p + \tau_z = \sum_m \frac{\gamma_m}{(\omega-\omega_m)^2+\gamma_m^2} + \sum_i \frac{\zeta_i}{(\omega-Z_i)^2+\zeta_i^2} = \sum_m \frac{\gamma_m^0+\gamma}{(\omega-\omega_m^0)^2+(\gamma_m^0+\gamma)^2} + \sum_i \frac{\zeta_i-\gamma}{(\omega-Z_i^0)^2+(\zeta_i^0-\gamma)^2}. \quad (11)$$

This is equal to the sum of spectral derivatives of the cumulative phases $d\theta_n/d\omega$ over all transmission eigenchannels. Since $\tau_T = \tau_p$ for a unitary system, $\tau_z$ must vanish. This can occur if zeros either lie on the real axis, $\zeta_i^0 = 0$, or are members of a conjugate pair with $\eta_{i\pm}^0 = Z_i^0 \pm i\zeta_i^0$. In the latter case, the two Lorentzian terms in $\tau_z$ associated with the pair have equal magnitudes but opposite signs so that their sum vanishes. This imposes topological constraints on the $\eta_i$ in a lossless system: As the system is deformed, any single zero must move along the real axis and any zero off the real axis must be part of a conjugate pair of zeros. When a system is deformed, the zeros of the conjugate pair either both move towards or away from the real axis. When a conjugate pair is brought to the real axis, the zeros meet at a ZP and are converted into two single zeros which subsequently move along the real axis. Any zeros on the real axis in lossless systems can only come off the real axis when it encounters another zero and the two zeros are converted into a conjugate pair that moves away from the real axis, as will be seen in Sec. IV.

When loss or gain are present, all zeros are displaced downward in the complex frequency plane by $-i\gamma$, breaking the reflection symmetry of $\eta_i$ relative to the real axis of the complex frequency plane. The contribution of a zero to the transmission time is then $\tau_{z_i} = \frac{\zeta_i^0-\gamma}{(\omega-Z_i^0)^2+(\zeta_i^0-\gamma)^2}$. The contribution of single zeros with $\zeta_i^0 = 0$ then no longer vanish and the contribution of a pair of transmission zeros at the same frequency with equal magnitude and opposite sign no longer cancel. Since the zero above the real axis of a pair moves closer to the real axis, while the zero below the real axis movers farther from the real axis, the contribution of the pair to $\tau_z$ is positive. However, once $\gamma > \zeta_i^0$, both zeros of the pair are in the lower half of the complex plane and the contributions of both Lorentzian to $\tau_z$ is negative. Since the integral over frequency of the Lorentzians line of a transmission zero is equal to $\pm\pi$ for a transmission zero above or below the real axis, and added absorption moves the zeros down in the complex plane, $\langle\tau_T\rangle$ is reduced by absorption.

We find in Fig. 3d that in random 1D samples, $\langle\tau_T\rangle = t_+ = \pi\langle\rho\rangle = \langle\tau_p\rangle$, so that $\langle\tau_z\rangle = 0$ even at extraordinarily high levels of absorption. But this can only be the case if transmission zeros do not exist or are further from the real axis than $\gamma$ for all values of $\gamma$ for which simulations were carried out in strictly 1D samples. In the argument that follows, we show that transmission zeros do not exist in strictly 1D samples. Assume for a moment that transmission does vanish in a 1D structure. Now consider the wave approaching the first interface after which the flux vanishes. Since the transmission coefficient of an interface which is not a perfect reflector does not vanish, a fraction of the flux impinging on the interface will be transmitted, giving nonvanishing transmission. But this contradicts the assumption that transmission vanishes in the sample. Thus, transmission cannot vanish in a strictly 1D structure. Reports of the vanishing of transmission in 1D structures with Fano line shapes [90,91] must therefore indicate that the samples studied are not strictly 1D. Similarly, calculations of nulls in conductance in systems containing coupled quantum dots with single channels leads [92,93] and in transmission through a cavity with a single incident and outgoing channel may arise because the interior of the system supports several internal transverse channels [37,94].

**D. Modal fit to simulated spectra in random 1D media**

In this section, we present a modal analysis of simulations of spectra of $\tau_T$, $T$, $E(x)$ and $U$ in a random 1D sample with and without loss or gain based on expressions developed in Sec. IIIC. It is not possible to demonstrate perfect agreement between the results of simulations and theoretical expressions in terms of modes since this would require the precise location of the unlimited number of poles in the complex frequency plane. We can, however, show that the modes are effectively complete in random 1D media by demonstrating that the fits to simulated spectra of propagation variables approach the results of simulations as the area in the complex plane in which the positions of poles are determined increases. The completeness of modes in quasi-1D random media, in which transmission zeros may exist, and the impact of transmission zeros on $\tau_T$ are discussed in Sec. IV.

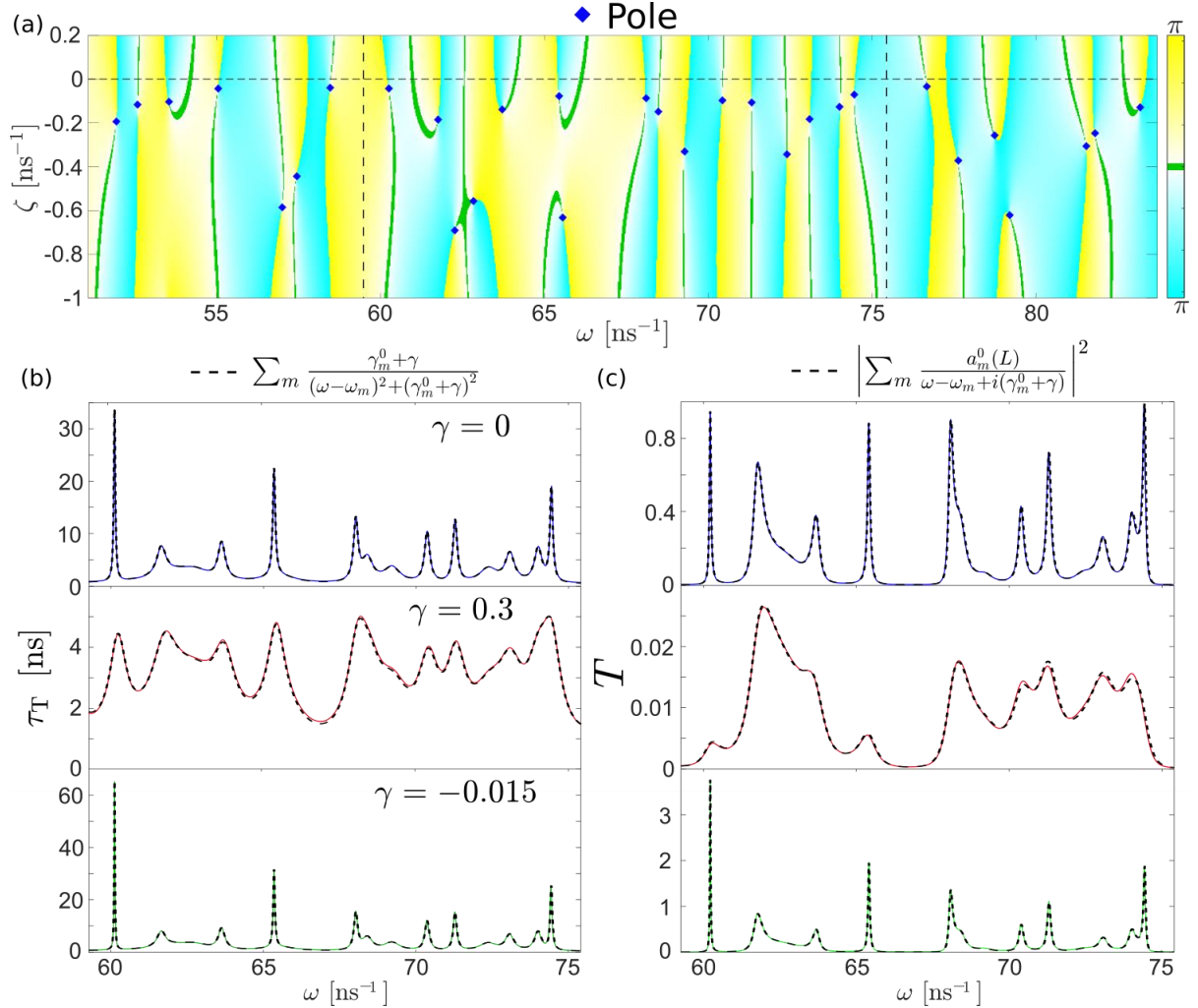

FIG. 5. Phase map and modal fits of spectra of transmission time and transmission. (a) A portion of the phase map of the transmitted field with poles indicated by blue diamonds. The poles over a wider range of $\zeta$ than shown in (a) are presented in Tab. I in Appendix H. (b) Simulations of transmission time without loss or gain ($\gamma = 0$ ns$^{-1}$, blue curves), with loss ($\gamma = 0.3$ ns$^{-1}$, red curves), with gain ($\gamma = -0.015$ ns$^{-1}$, green curves) and associated modal sums (black dashed curves) for the same sample of Fig. 4. (c) The corresponding transmission and associated modal sums.

The phase map and modal fits for $\tau_T$ and $T$ are shown in Fig. 5 for the same sample as in Fig. 4 with and without loss and gain. Simulations for quantities averaged over the random ensemble are presented in Fig. 3. The phase map of the transmitted field is shown in Fig. 5(a). The phase along lines parallel to the real axis in the imaginary frequency plane is found by calculating the phase of the transmitted field when uniform gain or loss are added. The poles indicated by blue diamonds are the singularities in the phase map. Since $\tau_z = 0$, we first consider the match of Eq. (11) to $\tau_T = \tau_p$ since this does not involve the amplitudes of the modes. A test of Eq. (11) is made by calculating the background that needs to be added to the sum in Eq. (11) due to modes in frequency windows of different widths. Alternatively, one can fit spectra of $\tau_T$ to a sum of modes within the spectral range and a slowly varying background function. In either approach, the magnitude of the background and the quality of the match to simulations is then evaluated as the spectral range in which the poles are determined is widened. The coordinates of the poles $(\omega_m^0, \gamma_m^0)$ in a larger area of the complex frequency plane for the 1D sample than is shown in Fig. 5(a) is given in Tab. I of Appendix I. The background is modelled as $a(\omega - \omega_{\min})^2 + b$. Since the poles are determined from the phase map, the only free parameters in the fit of Eq. (11) to simulations of $\tau_T$ using the poles within the spectral window are the three parameters of the background function, which represent the impact of modes which lie outside the spectral window. The solid curves in Figs. 5(b) and 6(a) are the results of the simulation, and the black dashed curves give the fits including the background. The background found from the fit for the unitary sample is presented as the dashed colored curves in Fig. 6(b) and seen to drop as the spectral widow is broadened with the value of the chi-squared test falling by more than two orders of magnitude, as given in the caption of Fig. 6. The differences between the results of simulations and the contribution of modes within different spectral windows to Eq. (11) is shown as the solid curves in Fig. 6(b) and seen to approximately track the background function found in the fit. The backgrounds within the central spectral windows due to modes outside spectral windows of different widths falls and flattens as the spectral window broadens.

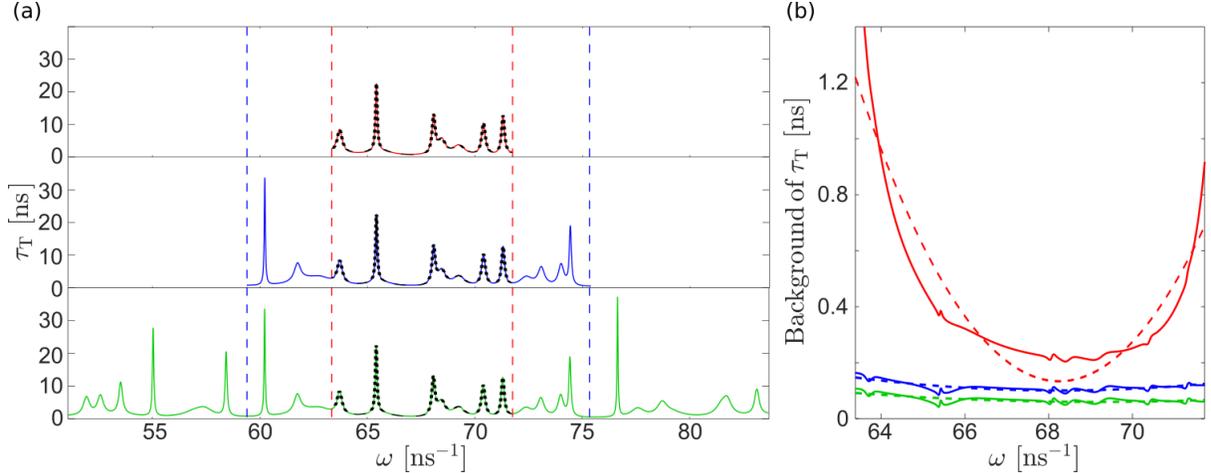

FIG. 6. Modal fit of transmission time in a unitary sample in spectral windows with different widths. (a) Simulations of $\tau_T$ in the sample of Fig. 5 (solid curves) and fits of the modal sum for $\tau_p$ in Eq. (11) with $\tau_z = 0$ plus a second order background (dashed curves). (b) Off-resonance quadratic background function found in fits in the central frequency region between dashed red vertical lines in (a). The background falls as the spectral window widens. The fit to $\tau_T$ in the central section of the spectrum between the vertical red dashed lines improves as the spectral window in which poles are determined modes increases with values of the chi-squared test $\frac{1}{N}\chi^2 = \frac{1}{N}\sum_{i=1}^{N}\frac{(O_i-E_i)^2}{E_i}$ of $4.83 \times 10^{-3}, 2.16 \times 10^{-5}$ and $2.09 \times 10^{-5}$ for the three spectra shown. The red solid line is the difference between the simulated spectrum of $\tau_T$ and the modal sum in the expression for $\tau_p$ in Eq. (11).

The transmitted field $E(L)$ and transmission $T \sim |E|^2$ in unitary and nonunitary samples are given by Eqs. (3) and (8). These variables depend upon the modal amplitudes of the transmitted field $a_m^0(L)$, as well as upon the poles. The modal amplitudes for the sample of Fig. 5(b) with $\gamma = 0$ at $x = L$ are found in the fit of the simulated spectrum of $E(L)$ by the expression in Eq. (3). The contribution to the field of off-resonance modes oscillates with frequency due to modal interference so we do not include a background in the fit. The values of $\omega_m^0$, $\gamma_m^0$, and $a_m^0(L)$ found for the unitary sample are used to obtain the black dashed curves for $T$ in the absorbing and amplifying samples shown in Figs. 5(c). Excellent agreement is found.

It is a general feature of simple mechanical and electromagnetic circuit resonances that resonances shift to lower frequencies in the presence of damping. Such a downward shift of modes in random media would correspond to an increase in the DOS with absorption. But the DOS is related to the wavelength which is not changed by absorption, as seen in the invariance of $\tau_T/t_+$ with absorption in Fig. 3(d), so that on average the central frequency of modes should not shift. The present results show that the modal frequency shifts are not observed up to the levels absorption of $\gamma = 0.3$ ns$^{-1}$.

We have seen that the modal fits to $E(L)$, which give the black dashed curves in Fig. 5(b) for $T$ are in excellent agreement with simulations for samples with and without loss and gain. We next consider the accuracy of the modal fit to the field throughout the sample and the completeness of modes in non-Hermitian systems.

In general, a wider spectral range is required to fit the field at points closer to the input surface. The quality of the fit to $E(x)$ is demonstrated in the plots of $|E(x)|^2$ with $x = 0.1L \sim 0.34\ell$ in Fig. 7(a) using modes over different frequency ranges in the lossless sample and in the plot in Fig. 7(b) in a strongly absorbing sample with $\gamma = 0.2$ ns$^{-1}$ in the widest frequency range. A plot of the in- and out-of-phase components of the field in the same configuration with and without the same loss as in Fig. 7 is shown Appendix F (Fig. 18). The fit in the narrowest spectral window shown between the dashed vertical lines in the top frame of Fig. 7(a) using only the modes in this region bears little resemblance to the simulated field. However, an excellent fit of Eq. (3a) to the simulated field in this central spectral region is achieved when the modes in the full spectrum shown are used in the fit of the central region, with a chi-squared test in the central sector reduced by three orders of magnitude relative to the fit using only modes in the central sector of the spectrum. Using the modal amplitudes and the known values of the poles found in the lossless sample of Fig. 7(a) and an absorption rate $\gamma = 0.2$ ns$^{-1}$ gives the black dashed curve in Fig 7(b), which overlaps the simulated spectrum shown as the red curve. The excellent agreement obtained even though the relative modal amplitudes are changed appreciably by absorption demonstrates the effective completeness of QNMs in random media when the spectrum is sufficiently wide.

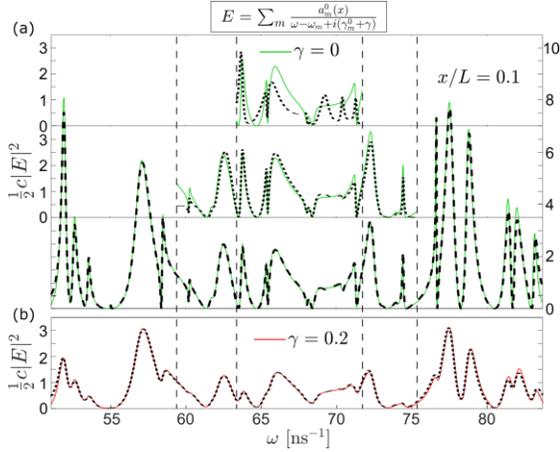

FIG. 7. Modal fits to the field inside the medium in spectral windows with different widths. (a) Simulations of $\frac{1}{2}c|E|^2$ (green curves) and the fit of the square magnitude of the field given in the coherent sum in Eq. (8) (black dashed curve) to spectra of increasing spectral width for the same sample as in Figs. 4 and 5 with $\gamma = 0$. The fit to $\frac{1}{2}c|E|^2$ in the center of the spectrum improves as the spectral window increases with values of the chi-squared test $\frac{1}{N}\chi^2 = \frac{1}{N}\sum_{i=1}^{N}\frac{(O_i-E_i)^2}{E_i}$ being $7.63 \times 10^{-3}, 2.85 \times 10^{-4}$ and $8.41 \times 10^{-6}$ for the three spectra shown. (b) The results for the widest spectrum for $\gamma = 0.2$ ns$^{-1}$. In the central sector of the spectrum, $\frac{1}{N}\chi^2 = 7.81 \times 10^{-4}$.

The excellent match of the modal sum to the simulated field for a point at a depth into the sample much smaller than the mean free path, at which the incident field has not yet been randomized, would suggest that a comparable fit could similarly be obtained for points still closer to the sample input. However, the quality of the fit to the square of the amplitude of the field within a given spectral window degrades rapidly as the position at which the field is evaluated moves closer to the input. This is seen in Fig. 19 of Appendix F for $x = 0.01\ell$. The fit improves with increasing spectral window but is poorer for each spectral window than for $x = 0.1\ell$ seen in Fig. 7. The larger spectral window required to obtain an equivalent quality of fit for smaller values of $x$ may be related to the shorter residence time of a fraction of the wave which is promptly scattered from the sample and is more strongly represented in the field at smaller depths into the sample.

We now study $U$ in terms of modes. In the absence of absorption, $U = 2\pi\rho$ so that $U$ is given by an incoherent sum of Lorentzians. We consider the impact of absorption on $U$ in samples with increasing modal overlap. We first consider the energy excited in a single isolated mode created at a quarter-wave defect state in the center of the band gap of the quarter wave stack shown in Fig. 8(a). The peak energy of the defect state shown in Fig. 8(b) is reduce by more than three orders of magnitude when loss at a level of $\gamma = 0.2$ is introduced, as seen in Fig. 8(c). At this level of absorption modes, at which modes of the pass band are broadened sufficiently that they overlap with the defect state, the peak intensity is reduced by less than 1% from the value predicted for a perfectly isolated mode of $U_m/2 = \frac{\gamma_m^0}{(\omega - \omega_m)^2 + \gamma_m^2}$. In contrast, simulations of $\tau_T$ are well fit by the incoherent sum over modes in Eq. (10) at all levels of absorption, as seen in Fig. 7d and in Fig. 4. Thus, the discrepancy is significant, and we might expect larger deviations when the degree of modal overlap increases.

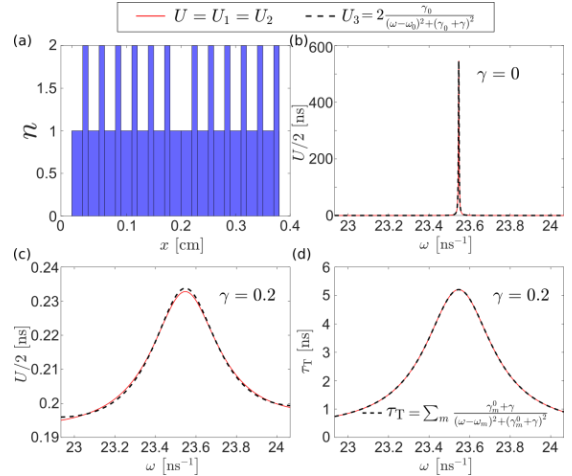

FIG. 8. Spectrum of total energy of near the center of the photonic band gap of a 1D photonic crystal with a single defect in the center of the sample. (a) The spatial distribution of refractive index of the sample. (b-d) Comparisons between simulations of spectra of total energy $U$ (red curve) and the incoherent modal sum of Eq. (9) $U_3$ (dashed black curve). The difference between these expressions increases with absorption. In part (c), the difference between curves is smaller than 1% even when the peak of the excited energy is reduced to 1% of its value in the lossless system.

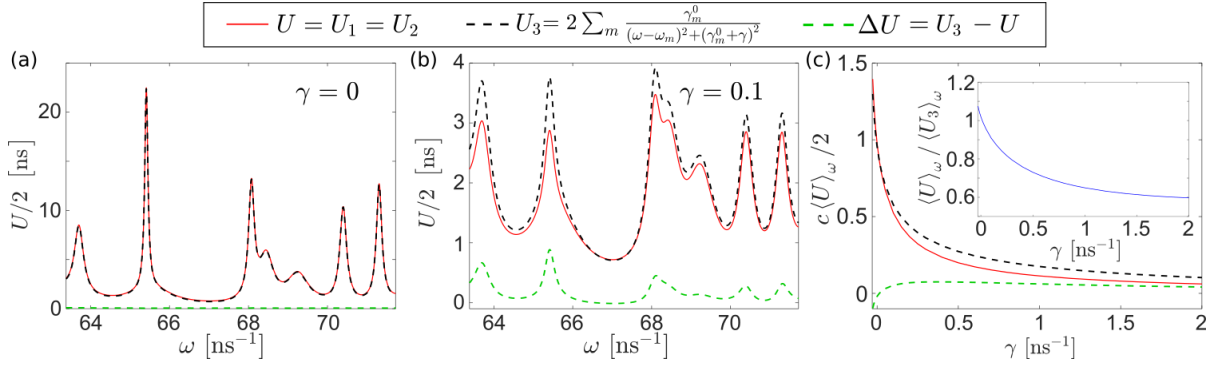

FIG. 9. Impact of loss and gain on total energy. Comparison of simulations of $U$ (red solid curves), the incoherent modal of Eq. (9b) (black dashed curves), $U_3$, and the difference $\Delta U = U_3 - U$ (green dashed curve) in a sample with (a) $\gamma = 0$, and (b) $\gamma = 0.1$ ns$^{-1}$. $U$ is lower/higher than the result of Eq. (9b) when absorption/gain is present. (c) Variation with $\gamma$ of $\langle U \rangle_\omega$ and $\langle U_3 \rangle_\omega$ and the difference $\langle \Delta U \rangle_\omega = \langle U_3 \rangle_\omega - \langle U \rangle_\omega$ for a single sample. $U$ falls below the incoherent sum in Eq. (9b) for $\gamma > 0$ and raised above it for $\gamma < 0$.

Simulations of the total energy $U$ in a unitary sample obtained from the scattered fields (Eq. (5)) and from the integral involving the square of the fields within the sample (Eq. (6)) are identical and are shown as the red curve in Fig. 9(a). This curve is seen to be in full agreement with the incoherent modal sum in Eq. (9) shown as the black dashed curve. Though Eqs. (5) and (6) give identical results for $U$ when absorption is present, these are no longer equivalent to the incoherent sum over modes in a nonunitary sample, with $\gamma_m^0 \to \gamma_m = \gamma_m^0 + \gamma$ in Eq. (9), $\sum_m \frac{\gamma_m^0}{(\omega - \omega_m)^2 + \gamma_m^2}$. The difference between the black dashed curve and the red curve is shown as the dashed green curve in Fig. 9(b) for a sample with $\gamma = 0.1$ ns$^{-1}$. This shows that peaks in simulations of spectra of $U$ are suppressed relative to the incoherent sum over modes. A plot of the variation with $\gamma$ of simulations of $U$ determined from Eq. (5) and of from the incoherent modal sum for the spectral window in Fig. 9(b) is shown in Fig. 9(c). The plots indicate that $U$ is reduced by absorption and enhanced by gain relative to the incoherent sum over modes.

Spectra of propagation variables in a single random configuration and the ensemble averages of these variables in random 1D samples for different values of $\gamma$ such that $\gamma_m^0 + \gamma > 0$ are seen in Figs. (1-8, 17-19) to be in full accord with Eqs. (5-8, 11) with values of $\omega_m^0$, $\gamma_m^0$, and $a_m^0(x)$ that are independent of $\gamma$. When the amplification rate within the medium equals the leakage rate from the medium for a particular mode, the modal linewidth vanishes and the energy inside the sample at resonance diverges. This corresponds to the modal lasing threshold. For stronger amplification, the effective linewidth becomes negative, $\gamma_m = \gamma_m^0 + \gamma = \gamma_m^0 - |\gamma| < 0$. We find that transfer-matrix simulations of spectra of $\tau_T, T$, $U$ and $\tau_D$ for the same spatial distribution of the real part of the index of refraction of the 1D sample studied thus far but with gain above the lasing threshold are fully in accord with Eqs. (7-9, 11), as seen in Fig. (20) of Appendix G. The transmission time, however, no longer corresponds to the DOS since the phase shift in tuning though a mode above the lasing threshold is $-\pi$ instead of $+\pi$, as it is in unitary or absorbing samples so that $\tau_T < \pi\rho$.

Previous transfer-matrix simulations in random 1D media and calculations based on the time-independent wave equation [95–97] noted a linear scaling of $\langle \ln T \rangle$ [98] well above the lasing threshold, $\langle \ln T(L) \rangle = -\frac{L}{\ell} - \frac{L}{|\ell_a|} + C$, where $|\ell_a| = -\ell_a$ is the gain length in an amplifying medium, as given in Eq. 11(b) in Appendix G. Here C is a positive constant well above the lasing threshold and equal to zero in absorbing samples. The dual symmetry of absorption and gain for systems below and above the lasing threshold with the same value $|\ell_a|$ runs counter to the known increase in energy density in lasing media that arises due to stimulated emission. The drop of transmission above the lasing threshold was shown to be an artifact of the use of the time independent wave equation, which disappears when time dependent simulations are used [99].

The source of the dual symmetry for $\langle \ln T(L) \rangle$ in the presence of loss and gain in random 1D media can be appreciated from a modal perspective. The numerators of the modal partial fractions at the sample in Eq. (8b) for $T(L)$, $a_m^0(L)$, as well as $\omega_m^0$ and $\gamma_m^0$ in the denominator are independent of $\gamma$. In the limit in which $|\gamma|$ is much greater than the half-linewidths of the modes in the interior, $\gamma_m = \gamma_m^0 + \gamma \to \gamma$ and $T \to \sum_{m,n} \frac{a_m^0(L)(a_n^0(L))^*}{[(\omega-\omega_m^0)-i\gamma][(\omega-\omega_n^0)+i\gamma]}$. Thus $T$ is the same for positive and negative $\gamma$. The dual symmetry of $\langle \ln T(\gamma) \rangle$ upon $\gamma$ shown in Fig. 21 of Appendix G is parallel to the previous finding of dual symmetry of $\langle \ln T(L) \rangle$ upon $L$ for absorption and gain.

### E. Measurements of spectra of transmission and transmission time

Measurements of field spectra in one sample from the ensemble of 100 random waveguides with $s = L/\ell = 3.39$ for which results were given in Fig. 2 are used to obtain the amplitude and phase of the transmitted field. Spectra of $\tau_T = d\varphi/d\omega$ and $T = |E(L)|^2$ are shown in Fig. 10 for this sample with the impact of the small natural absorption removed, (Fig. 10(a), $L/\ell_a = 0$), with absorption (Fig. 10(b), $L/\ell_a = 0.87$ or $\gamma = 0.12$), and with gain smaller than the leakage rate of the modes in the region (Fig. 10(c), $L/\ell_a = -0.05$ or $\gamma = -0.0065$). The number of modes with central frequency within the spectral window is determined from the total change in phase through the interval of $\Delta\varphi = 48$ giving $\Delta\varphi/\pi = 15$ modes. The plot of the phase in different spectra is smoother in samples with greater absorption but the total phase change, $\Delta\varphi$, is nearly the same.

The fit of $\tau_T$ for $\gamma = 0$ by the sum over modes within the frequency window of the measurements, and a background due to modes outside this frequency window is shown as the black dashed curve in Fig. 10(a). The values of $\omega_m^0$ and $\gamma_m^0$ found in the fit to spectra of $\tau_T$ in the unitary sample are used in the fit of spectra of the transmitted field to find the amplitudes of the transmitted field $a_m^0(L)$. Spectra of $T = |E(L)|^2$ are plotted as the black dashed curve in Fig. 10(a). Fits of comparable quality are found in samples with loss and gain obtained with $\omega_m = \omega_m^0$, $\gamma_m = \gamma_m^0 + \gamma$, and $a_m = a_m^0(L)$.

## IV. TRANSMISSION TIME IN MULTICHANNEL MEDIA

We now explore wave propagation in quasi-1D random samples in which transmission zeros may exist and contribute to $\tau_T$. We consider the impact of transmission zeros upon the transmission and transmission time of the entire system, $T$ and $\tau_T$ and upon individual transmission eigenchannels, $\tau_n$ and $t_n$. Since transmission zeros emerge from the interference of modal contributions to the field, the effective completeness of QNMs in quasi-1D media would be manifest in the close correspondence between simulations of $\tau_T$ and Eq. (11) when the transmission zeros and poles are known in a sufficiently large sector of the complex frequency plane. At the same time, the knowledge of the poles should yield spectra of transmission in accord with simulations.

The average of the lowest transmission eigenvalue for an $N$-channel conservative quasi-1D sample is exponentially small, with $\langle \tau_N \rangle \sim e^{-2L/\ell}$ [57,58,62,63]. Increasing the scattering strength might be expected to enhance the probability of creating transmission zeros near or on the real axis.

We first study the impact of absorption on the statistics of $\tau_T = \sum_{n=1}^N t_n$ in random rectangular samples coupled to empty waveguides on both sides with use of recursive Green's function simulations [98,99]. The random medium is shown schematically in Fig. 11(a). The samples composed of square elements with sides of length $a = \lambda_0/2\pi$, where $\lambda_0 = 1$ m is the vacuum wavelength, are coupled to waveguides supporting $N = 8$ channels. The sample width and length are $26a$ and $600a$, corresponding to 14.4 m and 95.5 m. The dielectric constant of each element is selected randomly from a rectangular distribution $[1 - \Delta\varepsilon, 1 + \Delta\varepsilon]$. The localization length is $1/\xi = \lim_{L\to\infty} \frac{d<\ln T>}{dL}$. Absorption is introduced by adding a constant imaginary part to the dielectric constant. The absorption lengths are determined from the decay length of energy in a homogeneous sample with loss, $1/\ell_a = \lim_{L\to\infty} \frac{d<\ln T>}{dL}$ and the corresponding absorption rate is $\gamma = \frac{1}{2}\frac{v_+}{\ell_a}$. $Nt_+ = \sum_{n=1}^N L/v_n = \frac{NL}{v_+} = \pi\langle\rho\rangle$.

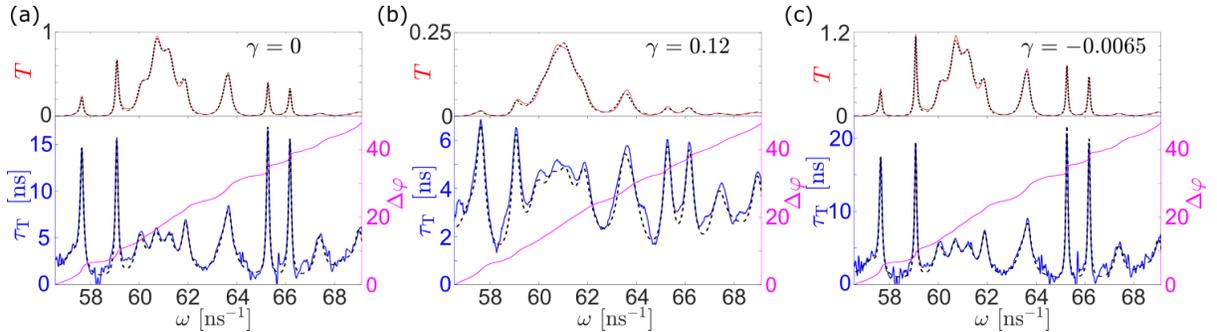

FIG. 10. Microwave measurements and modal fits of spectra of transmission, transmission time, and phase. (a-c) The transmission (red curve) and transmission time (blue curve) and fits of the associated coherent and incoherent modal sums (black dashed curve) in the legend for (a) the sample without absorption ($L/\ell_a = 0$), (b) the sample with loss ($L/\ell_a = 0.87$). and (c) the sample modified by adding gain ($L/\ell_a = -0.05$). Spectra of the phase are shown as purple curves.

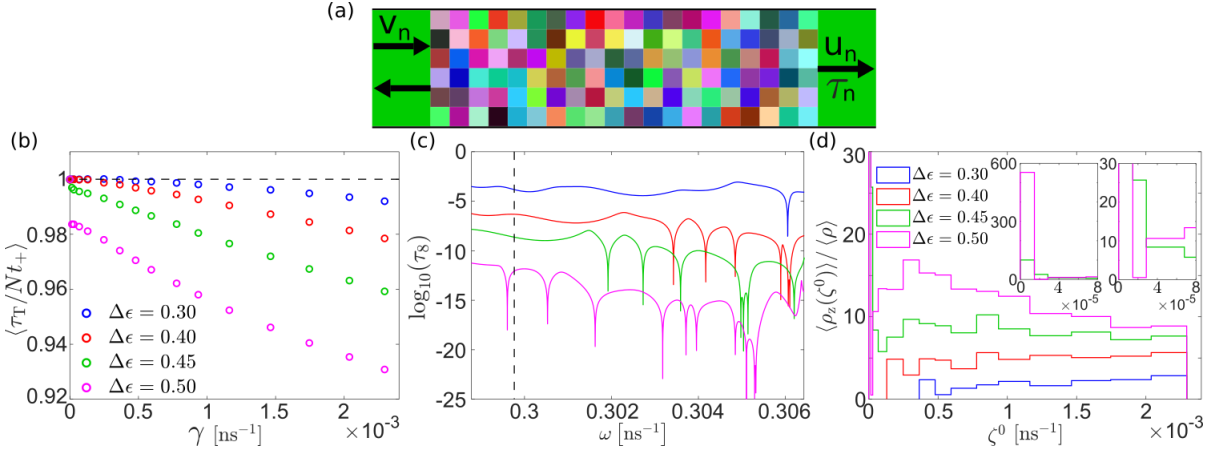

FIG. 11. Impact of absorption on the transmission time and the probability distribution functions of transmission zeros. (a) Schematic diagram of locally 2D quasi-1D system. Squares with different colors represent the dielectric constants in various square elements. The dielectric constant of the surrounding region is unity. (b) The normalized transmission time for different scattering strengths as a function of absorption. Results are given for $\Delta\varepsilon = 0.3$ (blue), $\Delta\varepsilon = 0.4$ (green), $\Delta\varepsilon = 0.45$ (red), and $\Delta\varepsilon = 0.50$ (purple), corresponding, respectively, to transmittance $g = 1.74, 0.98, 0.73, 0.54$ at the frequency $\omega = 0.2998$ ns$^{-1}$. (c) Spectra of the lowest transmission eigenvalue in a set of samples with different values of global disorder $\Delta\varepsilon$, which are the same as in (a) and all have the same spatial variation of the relative fluctuations in the local dielectric function from the average of unity, $(\varepsilon - 1)/\Delta\varepsilon$ in the range $[-1,1]$. Transmission zeros on the real axis are seen in the dips in spectra of $\log_{10}(\tau_8)$. The vertical dashed black line indicates the frequency at which results are obtained and analyzed in 11(b). (d) A portion of the probability distribution of the imaginary part of transmission zeros in a unitary sample for different scattering strengths calculated from the results in (b) using Eq. (12).

The combined effects of scattering and absorption upon $\langle\tau_T/Nt_+\rangle$ for samples are shown in Fig. 11(b). In addition to the results for the same value of $\Delta\epsilon = 0.3$ ($g = 1.74$, $L/\xi = 0.64$, blue circles) as in Fig. 10, results of simulations are shown for $\Delta\epsilon = 0.4$ ($g = 0.98$, $L/\xi = 1.20$, green circles); $\Delta\epsilon = 0.45$ ($g = 0.73$, $L/\xi = 1.65$, red circles); and $\Delta\epsilon = 0.5$ ($g = 0.54$, $L/\xi = 2.04$ purple circles). The number of absorption length $L/\ell_a$ ranges from 0 to 14.1, corresponding to the absorption rate $\gamma$ ranging from 0 to $2.23 \times 10^{-3}$ ns$^{-1}$. For the ensembles with the two smallest values of $\Delta\epsilon$, $\langle\tau_T\rangle$ is unchanged up to some value of $\gamma$ after which it falls gradually with added absorption. In contrast, for the ensembles with the two largest values of $\Delta\epsilon$, $\langle\tau_T\rangle$ drops below $Nt_+$ for the slightest additional absorption and then falls gradually. The magnitude of the discontinuity in $\langle\tau_T\rangle$ and the rate decay $\langle\tau_T\rangle$ with $\gamma$ generally increases with $\Delta\epsilon$.

Since $\langle\tau_p\rangle$ is the DOS, its contribution to $\langle\tau_T\rangle$ is independent of loss and scattering strength in quasi-1D media, as in 1D samples. However, $\langle\tau_z\rangle$ may fall with increasing absorption as transmission zeros are swept below the real axis as absorption increases. Single zeros on the real axis in a lossless sample are moved into the lower half of the complex plane for the slightest additional absorption and the upper transmission zero of a pair is brought below the real axis once $\gamma > \zeta_i^0$. Thus, the results in Fig. 11(b) indicate that at the frequency of the simulations there are no zeros on or very close to the real axis for the ensembles with $\Delta\epsilon$ equal to 0.3 and 0.4, while zeros do reside on and near the real axis for the ensembles with $\Delta\epsilon$ equal to 0.45 and 0.5.

Since transmission zeros are moved downward in the complex plane by $i\gamma$ when absorption is added, the average number of zeros at frequency $\omega$ brought below the real axis as $\gamma$ is incremented by $\Delta\gamma$ is proportional to the density of zeros in the corresponding lossless system with imaginary part $\zeta^0$, $\rho_z(\omega, \zeta^0 = \gamma)\Delta\gamma$. Further, since the contribution to $\tau_z$ of a single zero above/below the real axis is $\pi/-\pi$ when integrated over frequency, the corresponding change in $\tau_T$ due to a zero of a conjugate pair moving below the axis is $2\pi$. Since $\langle\tau_p\rangle$ is invariant with absorption, the change in the average transmission time is due solely to the transmission zeros, $\Delta\langle\tau_T\rangle = \Delta\langle\tau_z\rangle = -2\pi\langle\rho_z(\omega, \zeta^0 = \gamma)\rangle\Delta\gamma$. Normalizing to the average transmission time in the lossless systems of $\langle\tau_p\rangle = Nt_+ = \pi\langle\rho\rangle$, gives, $\frac{d\langle\tau_T/Nt_+\rangle}{d\gamma} = -2\langle\rho_z(\omega, \zeta^0 = \gamma)\rangle/\langle\rho\rangle$, for the contribution of conjugate pairs. When a real zero on the frequency axis in the lossless system is brought below the real axis, the change in the transmission time integrated over frequency is $-\pi$ since a real zero does not contribute to the transmission time. Letting $\rho_z^0(\omega,0)\delta(0)$ be the density of zeros on the real axis in the lossless system, gives,

$$\frac{d\langle \tau_T/Nt_+\rangle}{d\gamma} = -\frac{2\langle \rho_z(\omega,\zeta^0=\gamma)\rangle + \langle \rho_z^0(\omega)\rangle \delta(\zeta^0)}{\langle \rho \rangle}. \quad (12)$$

Thus, the continuous probability density of the imaginary part of zeros in the complex plane for unitary media in a lossless medium, $\langle \rho_z(\omega,\zeta^0)\rangle$, and the spectral density of zeros on the real axis $\langle \rho_z^0(\omega)\rangle$, at a frequency $\omega$ can be found from the sensitivity of $\langle \tau_T/Nt_+\rangle$ to absorption.

The presence of transmission zeros on the real axis can be seen in the vanishing of transmission of the lowest eigenchannel, $\tau_8$, in unitary samples. Spectra of $\log \tau_8$ in Fig. 11(c) for single configurations with the four values of $\Delta\epsilon$ as in Fig. 11(b) show that real transmission zeros move to lower frequencies as $\Delta\epsilon$ is increased and $g$ falls. The frequency at which $\langle \tau_T\rangle$ is determined in Fig. 11(b) is indicated as the vertical dashed black line in Fig. 11(c). For $\Delta\epsilon = 0.3$ and 0.4, Fig. 11(b) shows that $\langle \tau_T\rangle/Nt_+ = 1$ for small values of absorption so that there are no transmission zeros on or very near the real axis. Real zeros are observed at lower frequencies when disorder is increased. Thus, transmission zeros appear to move along the real axis to lower frequencies as $\Delta\epsilon$ increases. Transmission zeros are found at higher frequencies since the scattering from the square elements with sides smaller than the wavelength ($a = \lambda_0/2\pi$) increases with frequency.

The variation of $\langle \tau_T/Nt_+\rangle$ with $\gamma$, shown in Fig. 11(b), is used together with Eq. (12) to obtain the density of the imaginary part of zeros in the complex plane for unitary media of different scattering strengths shown in Fig. 11(d). The plots of $\rho_z(\zeta^0)$ in Fig. 11(d) show that in weakly scattering unitary samples there are no transmission zeros on or near the real axis. As $\Delta\epsilon$ increases, the likelihood of conjugate pairs of zeros being found near the real axis increases. With stronger scattering, zeros are found on the real axis and their number increases with scattering strength. Transmission zeros reach the real axis in a certain frequency range by either being created in this frequency range as the zeros in a conjugate pair merge on the real axis or by single zeros sliding along the real axis into the frequency range. Both of these processes are demonstrated in Fig. 13.

In addition to falling with increasing scattering strength or absorption, $\langle \tau_T\rangle$ falls as the number of channels increases for fixed scattering strength. The variation of $\langle \tau_T/Nt_+\rangle$ with $\gamma$ for the sample with $\Delta\epsilon = 0.4$ is shown in Fig. 12 for $N = 4, 8, 16$. For $N = 32$, $\langle \tau_T/Nt_+\rangle = 0.90$, but it is not possible to determine $\langle \tau_T/Nt_+\rangle$ for appreciable absorption since the lowest transmission eigenchannels fall rapidly with absorption. The curve for $N = 8$ corresponds to the red circles in Fig. 11(b). Increasing $N$ lowers the transmission time except at small values of $\gamma$ for $N = 4, 8$. Thus, transmission zeros in this sample appear on the real axis for higher values of $N$ even in the absence of absorption. Since transmission zeros are absent in 1D and tend more strongly to be on or near the real axis as $N$ increases, the increasing number of transmission zeros on the real axis may be associated with the increased transverse component of the $k$-vector near the sample output.

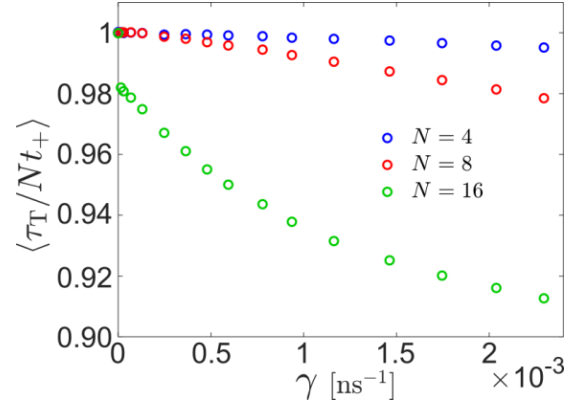

FIG. 12. The normalized transmission time for different channel numbers with same $\Delta\epsilon = 0.4$ as a function of absorption.

The distribution of transmission zeros in the complex frequency plane in random lossless quasi-1D samples and their evolution as the range of $\varepsilon$, $[1-\Delta\varepsilon, 1+\Delta\varepsilon]$, increases can be seen in the maps of the phase of $\det(t)$ for different values of $\Delta\varepsilon$. Figures 13(a-e) show maps of the phase of $\det(t)$ in samples in which the spatial distribution of relative fluctuations of the dielectric constant from its average of 1 is fixed, and only the range of fluctuation, $2\Delta\varepsilon$, changes. The relationship between $\det(t)$ and the zeros is given in Eq. (10). The zeros and poles of the TM are phase singularities, with properties similar to those of phase singularities in the speckle pattern of scattered light [100–104]. The phase changes by $-2\pi$ in a clockwise circuit around a zero and by $+2\pi$ around a pole. Using these properties of the phase singularities, we can identify the zeros and poles of the TM, which are indicated by red and blue diamonds, respectively, in Figs. 13(a-e).

Figure 13(a) shows four zeros and two poles. Two of the zeros form a conjugate pair and two zeros lie close together on the real axis. As $\Delta\varepsilon$ increases, the zeros of the conjugate pair move closer to the frequency axis (Fig. 13b), merge at a ZP (Fig. 13(c)) and two single zeros move away from the ZP on the real axis (Fig. 13(d) and 13(e)). The phase changes by $-4\pi$ in a clockwise rotation around the ZP in Fig. 13(c). The positions of the transmission zeros in the complex frequency plane is highly sensitive to changes in $\Delta\varepsilon$ near a ZP. This can be seen in the change of the position of transmission zeros between Fig. 13(b) and 13(c). A plot of the rapid change with $\Delta\varepsilon$ of the real and imaginary coordinates of the transmission zero near a ZP is seen in Fig. 13(f). Though transmission zeros are created when the zeros of a conjugate pair arrive at the real axis and by time reverse process can occur in which two single zeros merge on the real axis and are converted to two zeros of a conjugate pair that leave the real axis, Figs. 13(d) and 13(e) show that real transmission zeros can move through one another. Over the range of the change in $\Delta\varepsilon$ in Fig. 13, the positions of the poles shift slightly to lower frequencies. This is consistent with the counting number $N(\omega) \sim V/(\lambda/2)^2$ at the frequency of the poles in our sample, changing to second order with $\Delta\varepsilon$. Though the average of the dielectric constant remains 1 as $\Delta\varepsilon$ increases, the average of $N(\omega)$ at a fixed frequency would increase slightly because of the relatively small wavelength within elements with higher $\varepsilon$. Consequently, the poles move to lower frequency so that $N(\omega)$ at the frequency of a particular poles tends not to change.

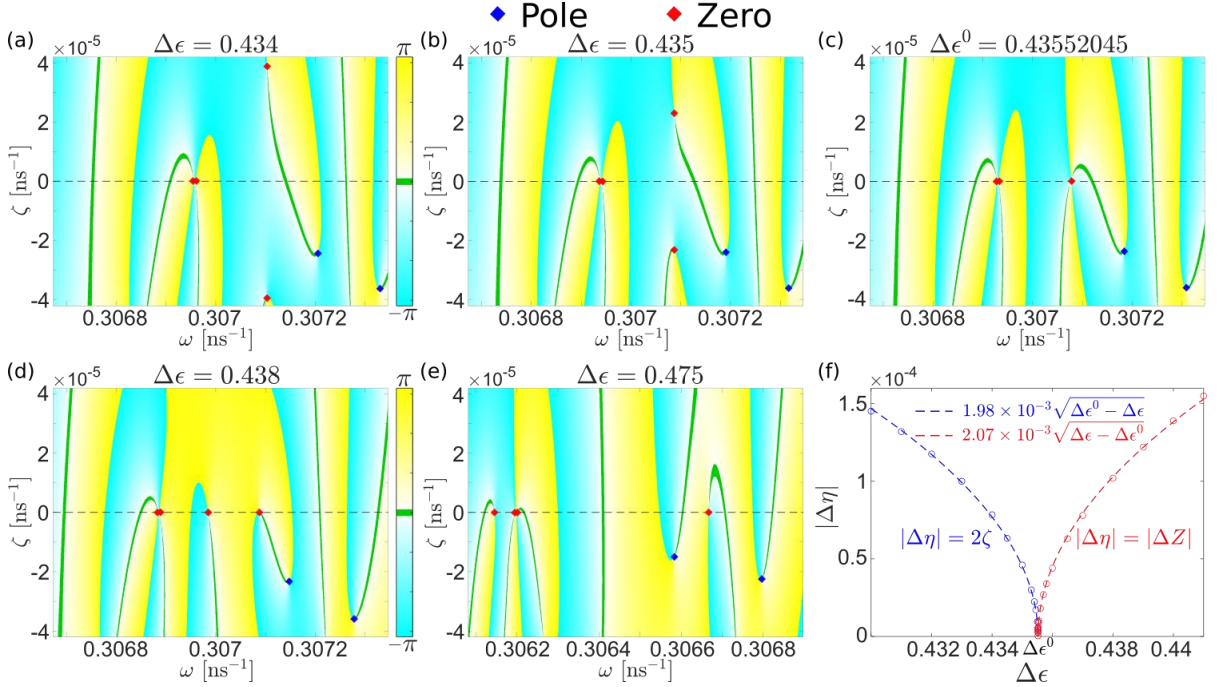

FIG. 13. Evolution of transmission zeros in samples with increasing disorder. (a-e) Phase maps of $\det(t)$ showing the motion of transmission zeros as the disorder characterized by $\Delta\varepsilon$ increases. The transmission zeros (red diamonds) and poles (blue diamonds) tend to shift to lower frequency as disorder increases. This necessitates a change in the frequency scale in (e). A conjugate pair of zeros in (a) approaches the real axis in (b) and converge at a ZP on the real axis in (c). The two single transmission zeros move away from the ZP in (d) and (e) as disorder increases. Two zeros that partially overlap on the real axis move together in (a-e). transmission zeros are highly sensitive to changes in the structure near a ZP, as compared to other transmission zeros or to the poles.

The difference in the coordinates of transmission zeros in the course of the conversion of a conjugate pair of zeros to two single zeros, which is presented in Figs. 13(a-e), is shown in Fig. 13(f) with fine steps in $\Delta\varepsilon$. The sensitivity to disorder, which can be expressed as the derivative of the spacing between two transmission zeros in the complex plane with respect to change of disorder, $\frac{d\Delta\eta}{d\Delta\varepsilon}$, is seen to diverge at the ZP [37]. The spacing between the conjugate pair of zeros along the imaginary axis in the complex plane before the zeros reach the ZP is $\Delta\eta = 2\zeta$. After the conjugate pair is transformed to two single zeros, the spacing on the real axis is $\Delta\eta = \Delta Z$. The separation between the zeros near the ZP shown in Fig. 13(f) is well fit by $\Delta\eta = \Delta\zeta = 2\zeta = 1.98 \times 10^{-3}\sqrt{\Delta\varepsilon - \Delta\varepsilon_{ZP}}$ for the conjugate pair, and $\Delta\eta = \Delta Z = 2.07 \times 10^{-3}\sqrt{\Delta\varepsilon - \Delta\varepsilon_{ZP}}$ for the single zeros on the frequency axis. This gives a square root singularity in the sensitivity of the separation between transmission zeros to $\Delta\varepsilon$. At the ZP, $\frac{d2\zeta}{d\Delta\varepsilon} = \frac{9.90 \times 10^{-4}}{\sqrt{\Delta\varepsilon - \Delta\varepsilon_{ZP}}}$ and $\frac{d\Delta Z}{d\Delta\varepsilon} = \frac{1.04 \times 10^{-3}}{\sqrt{\Delta\varepsilon - \Delta\varepsilon_{ZP}}}$, for the conjugate pair and the single zeros, respectively.

The high sensitivity of the imaginary parts of the transmission zeros of a conjugate pair of zeros near the frequency axis to changes in the structure indicates that the probability of finding transmission zeros with small values of $\zeta^0$ is small. A slight structural change either moves a transmission zero further from the real axis or towards the real axis where the conjugate pair of zeros is converted to two single zeros. As a result, $\rho_z(\zeta^0)$ is low for small values of $\zeta^0$ and rises as $\zeta^0$ increases, as can be seen in Fig. 11(d). Single zeros can accumulate on the real axis since they can only leave the real axis after each zero encounters another zero at a ZP and not every encounter of two single zeros results in the formation of a conjugate pair.

The impact of transmission zeros can be seen in a comparison of spectra of transmission and transmission time summed over all transmission eigenchannels, $T$ and $\tau_T$, and for individual eigenchannels, $\tau_n$ and $t_n$, with and without loss. In Fig. 14, we consider two samples with different global disorders of $\Delta\varepsilon = 0.435$ and 0.438 but with the same spatial distributions of the relative fluctuations of $(\varepsilon - 1)/\Delta\varepsilon$. The phase maps of $\det(t)$ for the respective samples with different global disorder are shown in Figs. 13(b) and 13(d). A conjugate pair of zeros is near the real axis of the first sample and two single transmission zeros created when the zeros of the conjugate pair met on the real axis at a ZP, as seen in Fig. 13(c), are on the real axis of the second sample. The spectrum of the phase of $\det(t)$ for in a random sample, such as those shown in Fig. 13, with a particular value of absorption corresponds to the phase variation along a line displaced by $i\gamma$ from the frequency axis. This is equivalent to the spectrum one would obtain on the real axis if the entire pattern were shifted down by $i\gamma$. Spectra in Fig. 14 for samples without loss are shown as solid curves, while spectra for samples with loss of $\gamma = 1.0 \times 10^{-5}$ are shown as dashed curves.

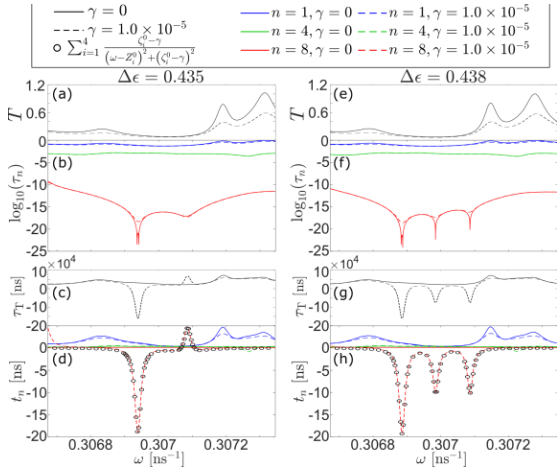

FIG. 14. Impact of transmission zeros on total and eigenchannel transmission and transmission time. Spectra of $T$ and $\tau_n$, and $\tau_T$ and $t_n$ without (solid curves) and with (dashed curves) absorption of $\gamma = 10^{-5}$ for $n$=1,4,8 in quasi-1D samples with $N = 8$. (a-d) Results for the sample with disorder $\Delta\varepsilon = 0.435$ with the phase map of det $(t)$ shown in Fig. 13(b). The sharp double dips in $\tau_8$ in the lossless system are due to two neighboring transmission zeros. These dips are washed out by the addition of a small loss of $\gamma = 1.0 \times 10^{-5}$. A conjugate pair of transmission zeros at $\omega = 0.307$ produces a mild dip in $\tau_8$. Linear plots of spectra of $\tau_8$ near a transmission zero are shown in the insets of (b) and (f). (d) When absorption is added, a single dip occurs in $t_8$ because the two transmission zeros overlap. A peak is produced as the upper transmission zero of the conjugate pair is moved closer to the real axis and the lower transmission zero is moved farther from the real axis. The curves of black circles in (d) and (h) are plots of Eq. (11), as explained in the text. (e-h) Results for the sample with $\Delta\varepsilon = 0.438$ with the phase map of det $(t)$ shown in Fig. 13(d). As in (b), the dips in $\tau_8$ seen in (f) are suppressed by small additional absorption. The dip in $t_8$ due to overlapping transmission zeros in the absorbing medium is nearly twice as deep as for the single transmission zeros because the frequency difference between the transmission zeros is less than the linewidths $\gamma$ of the transmission zeros.

In lossless samples, nulls occur in spectra of the lowest transmission eigenvalue $\tau_8$ at the frequencies of each of the real zeros, and peaks in $T$ and $\tau_8$ occur at the frequencies of poles lying close to the real axis. The two peaks in $T$ at the higher frequency end of each of the spectra are associated with the poles at the frequencies of these peak seen in Figs. 13(b) and 13(e). The peaks in transmission shift to lower frequency as $\Delta\varepsilon$ increases tracking the shift in the frequency of the poles. Spectra of the transmission time of the lowest transmission eigenchannel $t_8$ in Figs. 14(d) and 14(h) in lossless samples are flat because, on the one hand, the lowest transmission eigenchannel is a superposition of far-off-resonance modes for which the phase varies slowly [105], and on the other hand, $\tau_z$ vanishes.

When weak absorption is introduced, spectra of $T$ in Figs. 14(a) and 14(e) are slightly suppressed and nulls that appeared in $\tau_8$ in unitary samples become mild dips, as seen in Fig. 14(b) and 14(f). Dips appear in $\tau_T$ in Figs. 14(c) and 14(g) and in $t_8$ in Figs. 14(d) and 14(h) for each single transmission zero. The single dips in $\tau_T$ and in $t_8$ associated with the two closely spaced transmission zeros seen in Figs. 14(b) and 14(f) are nearly twice as deep as the dips associated with isolated zeros because the separation in frequency of these transmission zeros is much smaller than the half-width of the dips of each of these zeros of $\gamma$. Peaks appear in spectra of $\tau_T$ and $t_8$ in Figs. 14(c) and 14(d) for a conjugate pair of zeros due to the imbalance in the positive and negative contributions of the zeros of a conjugate pair produced when absorption is introduced. The transmission zero in the upper half of the complex plane moves closer to the real axis while the lower transmission zero is displaced farther from the real axis. This leads to a peak in the contribution of the conjugate pair to the transmission time, $\tau_{z_i} = \frac{\zeta_i^0 - \gamma}{(\omega - z_i^0)^2 + (\zeta_i^0 - \gamma)^2} +$ functions to spectra of $\tau_8$ near a single zero, double zero, and ZP in Fig. 14 are seen in Figs. 15(a-c).

In quasi-1D samples, the phase map of $\det(t)$ gives

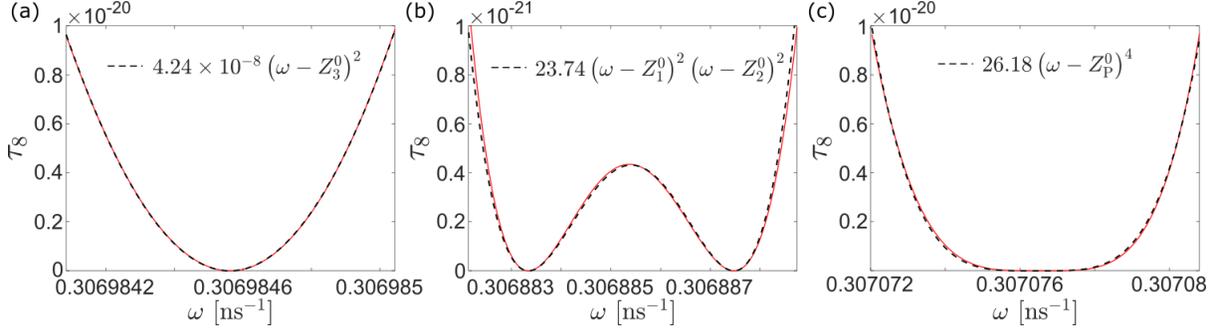

FIG. 15 Spectra of the lowest transmission eigenvalue near a transmission zero. The solid red curves are simulations of spectra of $\tau_8$ and the black dashed curves are fits using with the functional forms given in the figure which are forms of Eq. (10) near transmission zeros. Spectra for a sample with disorder strength $\Delta \varepsilon = 0.438$ near (a) for a single transmission zero and (b) for a pair of closely spaced transmission zeros, which can be seen in the phase map of Fig. 13(d). (c) The corresponding spectra near the ZP which is seen in Fig. 13(c).

$\frac{-\zeta_i^0 - \gamma}{(\omega - z_i^0)^2 + (-\zeta_i^0 - \gamma)^2}$, at $\omega = Z_i^0$.

The dips and peaks in $t_8$ are seen to be closely correlated with those in $\tau_T$. This suggests that the contribution of transmission zeros near the real axis to $\tau_T$ may be fully represented in $t_8$. We explore this conjecture by plotting the contribution of transmission zeros to $t_8$ plus a small flat background due to off-resonance modes that is not much disturbed by the introduction of moderate absorption. Simulations of $t_8$ for $\gamma = 10^{-5}$ shown as the dashed red curves in Figs. 14(d) and 14(h) and the sum of the expression for $\tau_z$ in Eq. (11) using the values of transmission zeros shown in Figs. 13(b) and 13(d) plus the constant background in $t_8$ in the lossless samples are shown as the curve of black ovals. The overlap of the two curves confirms that the contributions of transmission zeros to $\tau_T$ are fully represented in $t_8$.

Nulls in spectra of $\tau_8$ at each of the transmission zeros in the lossless sample are seen as sharp drops in $\log_{10}(\tau_8)$. Since transmission near a real transmission zero is small, the impact of the transmission zero should be manifest in the lowest transmission eigenchannel. Following Eq. (10), the transmission for this eigenchannel near a single real transmission zero at $\omega = Z_1$ should be $\tau_N = \tau_8 \sim (\omega - Z_1)^2$. This is confirmed in spectra of $\tau_8$ near transmission zeros that lie on the real axis shown in Fig. 15. The determinant of $t$, as given in Eq. (10), changes rapidly near real zeros on the real axis for which the factor $(\omega - \eta_i) = (\omega - Z_i)$ is real. For a single transmission zero, $\tau_8 \sim (\omega - Z_1)^2$, while for two closely spaced transmission zeros $\tau_8 \sim (\omega - Z_1)^2(\omega - Z_2)^2$. When two transmission zeros coincide at a ZP, $Z_1 = Z_2 = Z_{ZP}$, $\tau_N \sim (\omega - Z_{ZP})^4$. Excellent fits of these the poles and zeros and thus $\tau_T$ via Eq. (11). A portion of the phase map over the range of $\zeta^0 \in [-20, 5] \times 10^{-5}$ is shown in Fig. 16(a). The poles and transmission zeros over the range $[-100, 100] \times 10^{-5}$ are given in Tab. II in Appendix I. Simulations of $\tau_T$ in samples with different absorption in the narrower frequency range between the vertical black dashed lines in Fig. 16(a) are seen in Fig. 16(b) to be in excellent agreement with Eq. (11) using the singularities in the complex frequency plane in Tab. II. The effective completeness of the modes can be further demonstrated by showing that the transmitted flux between any pair of channels in the narrower frequency range can be found using the values of the singularities in Tab. II and the modal amplitudes as fitting parameters. An excellent fit to the spectrum of transmission between an exciting channel which is the in-phase linear combination of all incident channels with equal amplitude and the first waveguide mode at the output surface is demonstrated in Fig. 16(c) for different values of absorption. The agreement of $\tau_T$ with Eq. (11) and the quality of a modal fit to spectra of the field can never be perfect but can always be improved by increasing the region of the complex plane surrounding the probed region in which the singularities are determined.

Features in transmission spectra associated with single transmission zeros are generally broadened when absorption is increased. However, because of the presence of zeros in the upper half of the complex frequency plane, dips in transmission spectra and peaks or dips in the transmission time can be produced by bringing transmission zeros close to the real axis by adding absorption. This is demonstrated in Appendix H and shown in Fig. 22. As in weakly absorbing media, the impact of transmission zeros is manifested in $t_N$ in strongly absorbing samples but the background due to the poles is no longer small. Unlike single transmission zeros on the real axis that remain on the real axis as the sample structure is modified, the vanishing of transmission when the upper transmission zero of a conjugate pair of zeros is brought to the real axis by absorption is not robust to deformation of the sample.

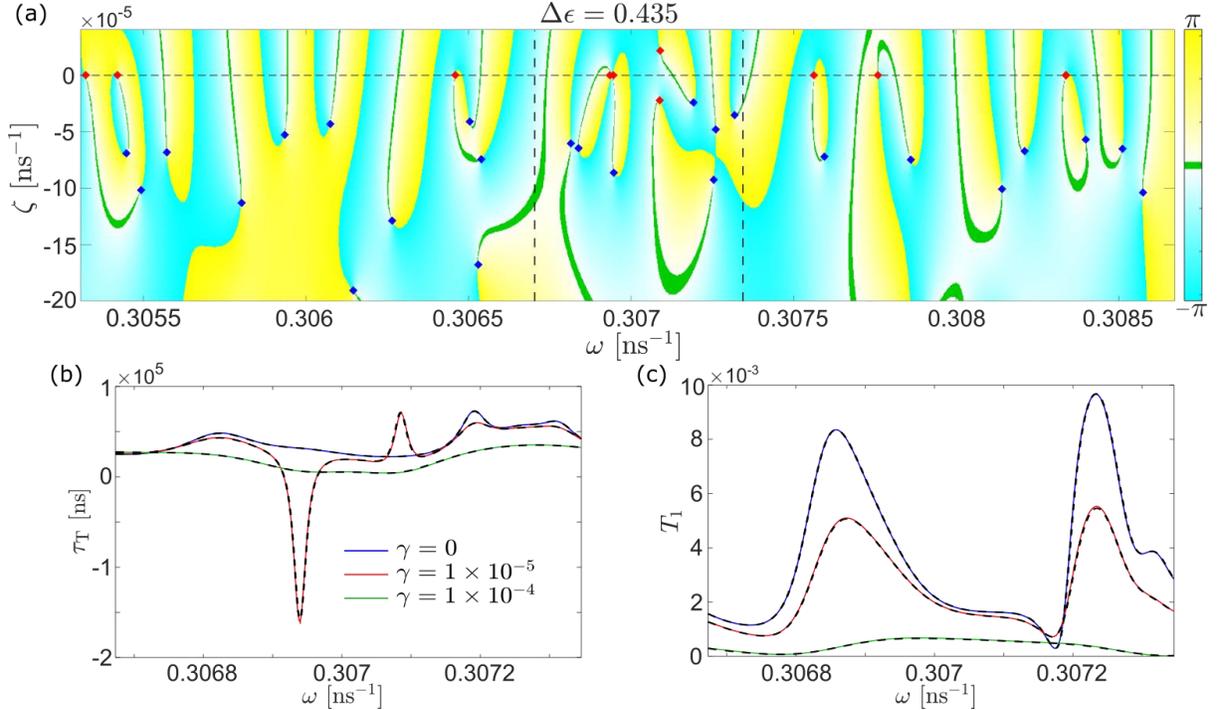

Fig. 16. Phase map and spectra of transmission time and transmission. (a) Phase map of $\det(t)$ over a wider range of the real and imaginary parts of the frequency than in Fig. 13. The phase map shown is a portion of the phase map used to determine the complex frequencies of the poles and transmission zeros used in the plots shown in (b) and (c). The poles and zeros over a wider range than shown in (a) with $\zeta$ up to $10^{-3}$ are presented in Tab. II in Appendix H. (c) Spectra of transmission time $\tau_T$. The blue, red and green solid curves show simulations of $\tau_T$ with, respectively, no absorption ($\gamma = 0$), weak absorption ($\gamma = 1 \times 10^{-5}$ ns$^{-1}$) and strong absorption ($\gamma = 1 \times 10^{-4}$ ns$^{-1}$). The dashed black curves are the sums of Lorentzian terms associated with modes and transmission zeros in Eq. (11) and a quadratic background. (c) Spectra of transmission of first waveguide channel. The square of the field transmission of incident channel $\mathbf{v}_{in} = \frac{1}{\sqrt{8}}[1,1,1,1,1,1,1,1]^T$ to the first waveguide mode $\mathbf{u}_1 = [1,0,0,0,0,0,0,0]^T$, $|t_{1,in}|^2$. The colored curves correspond to the same values of $\gamma$ as in (a). The dashed black curve is the fit of Eq. 8 to the simulations without a background.

We have seen that $\langle \tau_T \rangle$ is invariant with scattering strength and loss in 1D media but falls with scattering strength and loss in an $N = 8$ quasi-1D medium. This suggests that the presence of transmission zeros is associated with the existence of transverse flux within the medium that can only exist in systems supporting multiple channels in the bulk. Figure 12 shows a plot of $\langle \tau_T \rangle$ vs. $\gamma$ for an ensemble with $\Delta\varepsilon = 0.4$ for different numbers of channels. The reduction of $\langle \tau_T \rangle$ in the presence of absorption is greatest in samples with the largest number of channels and real zeros exist for $N = 16$. For $N = 16$, the decay rate of $\langle \tau_T \rangle$ with absorption falls as $\gamma$ increases. Thus, zeros exit in multichannel samples and their number on or near the real axis in unitary sample increases with $1/\ell$, $\gamma$, and $N$.

## V. CONCLUSION

An appreciation of the role of resonances in physical phenomena grew slowly in the centuries after Galileo's studies of freely oscillating pendulums. Euler drew the distinction between the driving frequency and the natural frequency of oscillation and Young showed that a pendulum could be made to oscillate at the driving frequency when a sinusoidal driving force is applied [108]. The study of resonances now extends far beyond the pendulum and the broader mechanical paradigm to encompass classical and quantum phenomena on all length scales.

When a medium without material loss is opened to its surroundings, modes are broadened as the poles are moved off the real axis into the lower half of the complex plane. The modes are then no longer orthogonal, but instead form a complete biorthogonal set [11,15]. In unitary samples, the Wigner time $\tau_W$, DOS $\rho$, dwell time $\tau_D$, total excited energy $U$ [8], and the transmission time, $\tau_T$ [54] are proportional to one another, as given in Eqs. (1-2). This work finds striking similarities and profound differences in the relationships of $E(x)$, $U$, $\tau_D$, and $\tau_T$ to the modes of the medium in unitary and dissipative 1D and quasi-1D media. This provides a deeper understanding of wave propagation in complex systems and points to open questions and potential applications.

The field excited within the lossless medium can be expressed as, $E(x) = \sum_m \frac{a_m^0(x)}{\omega - \omega_p}$, where $a_m^0(x)$ gives the spatial variation of the excited mode with a pole at $\omega_p = \omega_m^0 - i\gamma_m^0$. The field within a medium with uniform loss or gain may still be expressed as a coherent sum over modes with $E(x) = \sum_m \frac{a_m^0(x)}{\omega - \omega_p}$, where the coefficient of the modal partial fraction $a_m^0(x)$ is unchanged and $\omega_p = \omega_m^0 - i\gamma_m^0 - i\gamma$. This is an arresting demonstration of the completeness of modes. Unlike the excellent fit of Eq. (8a) to the simulated field in the lossless sample by finding the modal amplitudes once the poles are determined from the map of the phase of the transmitted field, there is no additional fitting when absorption is added. Though the relative weights of different modes change dramatically when loss is added, Eq. (8a) is still in excellent agreement with the simulated field. The modal expression for $E$ approaches the simulated field more closely as the spectral range of poles is increased in both 1D and quasi-1D systems. The spectral range required to give an excellent modal fit to the field grows gradually and then rapidly as the point of observation approaches the incident surface. This explains the difficulty of carrying out a modal fit in the interior of random media. It remains to carry out a full analysis of the quality of the modal fit to the field at different distances from the source.

The modal analysis might be expected to break down when modes are excited far-off-resonance since the half wavelength of the incident driving field differs from the correlation length in the random medium. We find, however, that modal expression for the field accurately match simulations even at high levels of dissipation, at which far-off resonance, and thus non-resonant excitation, is large. The impact of non-resonant contributions to the field might best be explored in small few-mode systems, which are of importance in photonics [23]. The methods developed here will be useful in such investigations.

When modal overlap in systems with loss or gain is weak, the total energy excited near a spectrally isolated mode is approximately given by the product of the contribution of the mode in the lossless medium and the factor $\gamma_m^0/(\gamma_m^0 + \gamma)$, as given in Eq. (9b). As modal overlap increases in an absorbing medium, $U$ is further reduced by destructive interference at each point in the medium. Though $U$ is no longer proportional to $\tau_W$ or $\tau_T$ in the presence of loss or gain, it remains equal to $\tau_D$. This highlights the strong connection between static and dynamic propagation.

We show that the average of the transmission time over an ensemble of random 1D samples is independent of the strengths of scattering, loss, ~~and~~ or gain. The transmission time, which is the derivative of the phase of the transmitted field, is then a sum of Lorentzian functions associated with the poles $\tau_p$ and is proportional to the DOS. In quasi-1D media, the transmission time is proportional to the DOS in lossless samples but is a sum of Lorentzian functions associated with both transmission zeros and poles, $\tau_T = \tau_p + \tau_z$, once loss or gain are added, as given in Eq. (11). The poles and zeros are singularities of the phase of $\det(t)$ in the complex frequency plane. The vanishing of $\tau_z$ in lossless multichannel systems imposes topological constraints on the position and motion of transmission zeros in the complex frequency plane under deformation of the medium: transmission zeros either lie on the real axis or appear as complex conjugate pairs. When uniform absorption or gain are present, transmission zeros move down or up by $i\gamma$ in the complex plane. This leads to a reduction in $\langle\tau_T\rangle$ with absorption as transmission zeros are swept below the real axis. The probability distribution of transmission zeros in the complex plane of the unitary system may be determined from the variation of the average transmission time with $\gamma$. The distribution of transmission zeros is expected to differ from the distributions for zeros of the scattering and reflection matrices since the transmission zeros do not correspond to eigenvalues of an effective internal Hamiltonian. The zeros of the SM are associated with the poles and are eigenvalues of an effective Hamiltonian in which all channels provide effective gain, whereas the zeros of the RM correspond to an effective Hamiltonian in which the input channel act as effective gain and the output channels act as effective loss [34,36,37].

As the structure of a unitary sample is modified, a conjugate pair of transmission zeros and two single zeros on the real axis may interconvert at a ZP. All real transmission zeros produce sharp dips in transmission that are sensitive to changes in the structure. The change in the frequency difference between two real transmission zeros has a square-root singularity near a ZP. The frequency difference of transmission nulls near a ZP with changes in the sample is therefore ultrasensitive to changes in the sample. Because the frequency difference between two transmission zeros can be determined directly from measurements of transmission in a sample without loss or of the transmission time in a system with small dissipation, it is not necessary to determine the shift in the absolute frequencies of the transmission zeros.

We note that the difference between 1D and quasi-1D systems regarding the presence of transmission zeros is not the number of channels coupling to the sample but the transverse dimensions of the scattering region. Transmission zeros may exist for conductors and cavities with a single input and output channel [92,93,37], but the transverse dimensions of the scattering region must be large enough to support more than a single transverse mode. In such systems, the wave near the output channel does not couple to the channel and the flux flows past the output channel. The drop in the transmission time for a medium of given mean free path and absorption but larger channel number is consistent with the association of vanishing transmission with transverse flux within the medium. Since the sensitivity is related to the phase change in transmission, greater sensitivity to small structural variations is achieved for shorter wavelengths. Optical measurements could probe deep-subwavelength changes in a structure. A square-root singularity with sample change also arises for the spacing between poles at an exceptional point (EP) at which distinct modes coalesce and have identical modal shapes [109,110]. This enables sensitive detection of structural change. In contrast to EPs, ZPs do not require that modes coalesce, and they can always be created by perturbing the medium.

The map of the phase of $\det(t)$ gives the poles and zeros of the TM that determine the transmission time in the medium at any level of uniform loss or gain. The poles and zeros for any level of loss or gain are obtained from the phase map in the unitary system by shifting the phase map down or up by $i\gamma$, respectively. In unitary samples, $\tau_N$ vanishes when transmission zeros appear on the real axis, so that the dynamic range of transmission eigenvalues is not only exponentially large [57] but diverges. In contrast, spectra of $t_N$ are flat in unitary samples since $\tau_z = 0$ while the lowest transmission eigenchannel in samples without absorption is due to modes that are far off resonance [107]. When small uniform absorption is present, however, the nulls in the lowest transmission eigenchannel are washed out while sharp dips or peaks are produced in $\tau_T$. Dips in $\tau_T$ are produced at the frequencies of single transmission zeros that are moved slightly below the real axis while peaks are produced at the frequencies of pairs of transmission zeros lying close to the real axis as the upper member of the pair is moved closer to and the lower member is move farther from the real axis. The spectrum of $t_N$ is then the sum of a small flat background and $\tau_z$. Transmission may also vanish in absorbing media when the upper transmission zero of a conjugate pair is brought to the real axis in the complex frequency plane. Sharp spectra in $t_N$ will then appear when the transmission is displaced slightly from the real axis.

The impact of absorption is of great importance in studies of Anderson localization of classical waves. Anderson [81] showed that a transition between diffusive and localized transport of electrons and electronic spins can arise with increasing disorder in 3D lattices and Gertsenshtein and Vasil'ev [111] demonstrated that electromagnetic radiation could be localized in single-mode waveguides even by weak disorder. The dephasing of electrons in solids by phonon scattering has restricted studies of coherent electron transport to mesoscopic samples between the nanometer and micron scales at ultralow temperatures [59,112,113]. In contrast, thermal fluctuations do not lead to dephasing of electromagnetic or acoustic waves even in macroscopic samples since the scale of such fluctuations are so much smaller than the wavelength of classical waves. This opens up the study of mesoscopic propagation of classical waves even for macroscopic samples [111,114–120]. Classical localization is of interest because it always occurs in large enough random systems with dimensionality of two or lower and because of the possibility of studying a pure Anderson transition in 3D without particles of the wave interacting via the Coulomb interaction or being trapped in a potential well [114]. However, the search for photon localization in 3D has been complicated by absorption, whose influence is particularly acute in media supporting long-lived localized states, and by fluorescence.

In the presence of absorption, total transmission in diffusive 3D samples falls inversely with thickness up to the absorption decay length $L_{\text{abs}} = \sqrt{\ell \ell_a/3}$, but then falls exponentially with this decay length [118]. The signature exponential falloff of transmission expected for localized waves in lossless media is hard to distinguish from the exponentially decay that occurs in absorbing diffusive media [114,121–124]. The decay of long-lived localized modes is seen in pulsed transmission in low dimensional samples [21], but fluorescence can complicate the interpretation of optical measurements in the time domain in random 3D slabs [124,125]. It has been shown that electromagnetic localization does not occur in random collections of point dipoles in random 3D media once polarization effects are included [126]. Whether photons can be localized in random dielectric 3D media and how this can be determined experimentally remains an open question.

The decreased spatial extent of excitation into a medium when absorption is added is often described as a shortening of the localization length. Thus, the localization length in 1D is taken to be the inverse of $-\langle \ln T \rangle$, which is reduced from $\ell$ in a lossless sample to $\left(\frac{1}{\ell} + \frac{1}{\ell_a}\right)^{-1}$ when absorption is added [95]. We have seen, however, that $a_m(x) = a_m^0(x)$ so that the shape of localized modes and the leakage rate through the boundaries is not changed by uniform absorption. Thus, the degree of modal localization is independent of absorption. The Thouless number $\delta = \delta\omega/\Delta\omega$ is a key localization parameter in lossless media because it relates spatial localization to spectral properties of modes, with $\delta = g$. In unitary media, $\delta\omega = \langle\gamma_m^0\rangle$ so that $\delta = \langle\gamma_m^0\rangle\langle\rho\rangle$. In the presence of dissipation, $\langle\rho\rangle$ does not change, but modes broaden so that $\delta\omega/\Delta\omega$ increases. On the other hand, $g$ is suppressed by absorption so that the possibility of localization cannot be given strictly in terms of these localization parameters. The challenge is thus to find the scaling of $\delta$ in equivalent lossless samples from measurements in samples with loss. This is as true for the propagation of ultrasound, where localization has been observed in 3D structures [119,120], as it is for electromagnetic localization,

In lower dimensional 1D, 2D and quasi-1D geometries, waves may be localized in small enough samples that the level spacing is large enough relative to the leakage and absorption rates that resonances can be discerned in transmission. It is then possible to determine the central frequencies and linewidths of the modes from transmission spectra and to determine $\langle\rho\rangle$ and $\langle\gamma_m\rangle$ for an ensemble of random samples [21,22]. The field absorption rate could be determined from the narrowed linewidth of transmission peaks in samples with reflectors placed over the boundaries. This would give $\langle\gamma_m^0\rangle = \langle\gamma_m\rangle - \gamma$ so that $\delta = \langle\gamma_m^0\rangle\langle\rho\rangle$ can be determined. But in random 3D media, the DOS rises rapidly as the sample size increases and absorption may wash out peaks in total transmission.

We now consider an approach to finding the scaling of $\delta$ from optical measurements of $\langle U\rangle$ and $\langle\tau_\mathrm{T}\rangle$ in random absorbing wedges. The determination of $\delta$ from $\langle U\rangle$ and $\langle\tau_\mathrm{T}\rangle$ would be relatively straightforward if these variables could be expressed as a sum of modal terms. But, in the presence of absorption, $\langle U\rangle$ may be lowered by destructive interference among overlapping modes and $\langle\tau_\mathrm{T}\rangle$ by transmission zeros that are lowered into the lower half of the complex plane. These factors are studied in Figs. 9, 11, and 12, and could be fully characterized in a fuller study of the impact of absorption in samples of different scattering strength and number of channels. Here, we illustrate the approach without taking these factors into account. The ratio of average energy and average transmission time is then $\frac{\langle U\rangle}{\langle\tau_\mathrm{T}\rangle} \sim \frac{\langle\sum_m \frac{\gamma_m^0}{(\omega-\omega_m)^2+\gamma_m^2}\rangle}{\langle\sum_m \frac{\gamma_m}{(\omega-\omega_m)^2+\gamma_m^2}\rangle}$, which can be evaluated by first integrating over frequency and then averaging over configuration to give, $\frac{\langle U\rangle}{\langle\tau_\mathrm{T}\rangle} = \frac{\langle\sum_m \gamma_m^0\rangle}{\langle\sum_m \gamma_m\rangle} = \frac{\langle\gamma_m^0\rangle}{\langle\gamma_m^0+\gamma\rangle}$. $\langle U\rangle$ could be determined from a comparison of optical measurements of total transmission and reflection in the sample under study, as given in Eq. (5), to measurements in diffusive samples with negligible absorption, for which $L < L_\mathrm{abs}$. The transmission time could be found from the determination of the contribution of the sample to measurements of the average spectral derivative of the phase in the far-field speckle pattern and an evaluation of the effective number of channels $N$ versus $L$, $\langle\tau_\mathrm{T}\rangle \sim N\langle d\varphi/d\omega\rangle \sim \pi\rho$. With an estimate for $\gamma$ and fuller knowledge of the impact of interference on $\langle U\rangle$ and of transmission on $\langle\tau_\mathrm{T}\rangle$ than given in this preliminary study, the scaling of $\frac{\langle U\rangle}{\langle\tau_\mathrm{T}\rangle}$, would give the scaling of the Thouless number in the equivalent lossless slab.

In this work, we have considered dispersionless media with uniform loss or gain in which the TM is fully determined. When fluctuations of the absorption rate are on a scale smaller than the spatial extent of a mode, the effect may be expected to be small, but nonuniformity in dissipation on a larger scale can lead to hopping of energy between modes [127–130]. Dispersion may alter the frequency of modes, but the modal description given here does not depend on dispersion in the dielectric constant and should not change. Incompleteness the TM reduces the dynamic range in the average of transmission eigenvalues [63,131], but transmission zeros still occur in multichannel systems. However, the zeros are no longer topological: single zeros can appear off the real axis and conjugate zeros do not occur.

Among the statistical questions regarding transmission zeros in unitary systems are the distribution of the real and imaginary parts of the spacing between transmission zeros in different frequency ranges for different strengths of disorder, and possible correlation between transmission zeros and between transmission zeros and poles. This is related to the motion of transmission zeros. For example, the zeros of a conjugate pair near the real axis move quickly with sample change heading towards or away from a ZP and so the probability of finding such zeros is low. Zeros on the real axis can move through one another and two closely spaced real transmission zeros can move together as the sample is changed as seen in Fig. 13. The physical basis for the different types of motion requires further investigation.

This work extends the paradigm of resonances to non-Hermitian system and find new relations and phenomena linking the excited field and energy, transmission time, and the density of states. Wave propagation in non-Hermitian systems is explained via the poles and zeros which appear as singularities in the phase map of the determinant of the transmission matrix in the complex energy or frequency plane. The analysis uncovers striking differences and similarities between waves in samples with and without loss or gain, and in the 1D and quasi-1D geometries. The field in a particular nonunitary medium is given in terms of the same parameters as in lossless media except that the modal linewidth becomes $\gamma_m = \gamma_m^0 + \gamma$ corresponding to a shift of the poles by $i\gamma$. When uniform loss is added to a sample, the total energy excited within the sample is no longer equal to the transmission time but remains equal to the dwell time. For small spectral overlap of modes, the energy remains an incoherent sum over modes, but with each modal contribution multiplied by a factor of $\gamma_m^0/(\gamma_m^0 + \gamma)$. When modes overlap, however, the energy excited is further suppressed by loss or enhanced by gain. In 1D, the transmission time is always proportional to the DOS and its ensemble average is independent of scattering strength, loss and gain. Transmission in 1D can never vanish, and the transmission time is equal to the sum of Lorentzian functions associated with the poles. In multichannel media, however, the transmission time in nonunitary media is suppressed by scattering and loss and enhanced by gain and becomes a sum over Lorentzians associated with transmission zeros as well as with poles. Sharp structure in spectra of transmission and transmission time may be created by displacing the transmission zeros by modifying the structure of a sample or adding loss or gain. This provides a path to ultrasensitive detection and narrowband filtering. Since the analysis is in terms of the poles and zeros of the medium without reference to the structure, the results are broadly applicable to complex systems.

## ACKNOWLEDGMENT

This work is supported by the National Science Foundation under EAGER Award No. 2022629 and by PSC-CUNY under Award No. 63822-00 51.

## APPENDIX A: DESIGN OF 1D LAYERED MEDIUM WITH DESIRED MEAN FREE PATH

We find the values of the indices of refraction, $n_{1,2} = 1 \pm \Delta n$ of a binary layered medium with $N_{\text{layer}}$ layers that produce a desired value of $L/\ell$. In a random 1D medium the average of the logarithm of transmission is the sum of the corresponding averages for the segments and given by, $\langle \ln T \rangle = \sum_1^{N_{\text{layer}}} \langle \ln T_i \rangle = -L/\ell$ [73]. In a binary system with alternating layer, the transmission coefficient of a normally incident wave is $\left(\frac{2n_1}{n_1+n_2}\right)^2$ (layer 1 to layer 2) and $\left(\frac{2n_2}{n_1+n_2}\right)^2$ (from layer 2 to layer 1). Thus for an ensemble with random spacing between interfaces, and $N_{\text{layer}} \gg 1$, $<\ln T> = -\frac{L}{\ell} = 2\ln\left[\frac{2n_1}{n_1+n_2}\frac{2n_2}{n_1+n_2}\right]^{N_{\text{layer}}/2} = N_{\text{layer}}\ln\frac{4n_1n_2}{(n_1+n_2)^2}$. Here we have ignored the difference in scattering at the sample boundaries. Since $n_1 + n_2 = 2$, this gives, $\frac{L}{\ell} = -N_{\text{layer}} \ln[n_1(2-n_1)]$. This method gives a value of $\ell$ which is within 0.3% of the results obtained from a comparison of simulations to $<\ln u(x)v_+> = -x/\ell$ [85]. The discrepancy is due to the weaker scattering at the first and last interfaces in the sample due to the smaller index mismatch than for interfaces in the interior.

## APPENDIX B: REMOVING IMPACT OF ABSORPTION IN MICROWAVE MEASUREMENTS

The average absorption rate in the medium, $\gamma$, is found from the half linewidths of long-lived modes peaked in the middle of a sample for which the leakage rate from the boundaries is small. The leakage rate is further reduced by placing a totally reflecting aluminum block at the outgoing boundary and a 90% reflector at the incident boundary, which allows a fraction of the energy launched in the waveguide to enter the sample. The linewidth of the narrowest modes, $\gamma = 4.7 \times 10^{-3}$ ns$^{-1}$, is then essentially the absorption rate of energy. The impact of absorption is compensated for by multiplying the spectrum by a broad Gaussian function and Fourier transforming from the frequency to the time domain to give the response to a narrow Gaussian pulse in time. The resulting variation of the field in time is then multiplied by the factor $e^{\gamma t}$ to compensate for loss due to absorption. Finally, the field is Fourier transformed back to the frequency domain to give the spectrum that would be given without loss. The validity of this approach is confirmed by the finding that the probability distribution function of intensity found in an absorbing random quasi-1D sample in which the impact of absorption is removed is in agreement with that predicted theoretically [82,119,132].

## APPENDIX C: WIGNER TIME DELAY IN 1D UNITARY AND NON-UNITARY SYSTEMS

The Wigner time delay is equal to the DOS in unitary media but becomes complex in the presence of loss or gain. Its real part is plotted in Fig. 3(d). The Wigner time delay is defined as $\tau_W = \text{Tr}\left(-iS^+ \frac{dS}{d\omega}\right)$, with $S = \begin{pmatrix} t & r' \\ r & t' \end{pmatrix}$. In a 1D-system, $\text{Tr}\left(S^+ \frac{dS}{d\omega}\right) = t^* \frac{dt}{d\omega} + t'^* \frac{dt'}{d\omega} + r^* \frac{dr}{d\omega} + r'^* \frac{dr'}{d\omega} = \frac{1}{2}\frac{d|t|^2}{d\omega} + i|t|^2 \frac{d\varphi_t}{d\omega} + \frac{1}{2}\frac{d|t'|^2}{d\omega} + i|t'|^2 \frac{d\varphi_{t'}}{d\omega} + \frac{1}{2}\frac{d|r|^2}{d\omega} + i|r|^2 \frac{d\varphi_r}{d\omega} + \frac{1}{2}\frac{d|C|^2}{d\omega} + i|r'|^2 \frac{d\varphi_{r'}}{d\omega} = \frac{1}{2}\frac{d}{d\omega}(|t|^2 + |t'|^2 + |r|^2 + |r'|^2) + i\left(|t|^2 \frac{d\varphi_t}{d\omega} + |t'|^2 \frac{d\varphi_{t'}}{d\omega} + |r|^2 \frac{d\varphi_r}{d\omega} + |r'|^2 \frac{d\varphi_{r'}}{d\omega}\right)$

For a unitary sample, $t = t', |r| = |r'|, |t|^2 + |r|^2 =$

$1, 2\varphi_t = \varphi_r + \varphi_{r'} + \pi$, so $\text{Tr}\left(S^+ \frac{dS}{d\omega}\right) = \frac{1}{2}\frac{d}{d\omega}(2) + i\left[2|t|^2 \frac{d\varphi_t}{d\omega} + |r|^2 \frac{d}{d\omega}(\varphi_r + \varphi_{r'})\right] = 0 + i\left[2|t|^2 \frac{d\varphi_t}{d\omega} + 2|r|^2 \frac{d\varphi_t}{d\omega}\right] = i2\frac{d\varphi_t}{d\omega}$. This gives $\tau_W = -i\text{Tr}\left(S^+ \frac{dS}{d\omega}\right) = 2\frac{d\varphi_t}{d\omega}$.

In a system with loss, $t = t'$ but $|r| \neq |r'|$ and $|t|^2 + |r|^2 \neq 2, 2\varphi_t \neq \varphi_r + \varphi_{r'} + \pi$. Now, $\text{Tr}\left(S^+ \frac{dS}{d\omega}\right) = \frac{1}{2}\frac{d}{d\omega}(2|t|^2 + |r|^2 + |r'|^2) + i\left(2|t|^2 \frac{d\varphi_t}{d\omega} + |r|^2 \frac{d\varphi_r}{d\omega} + |r'|^2 \frac{d\varphi_{r'}}{d\omega}\right)$, so that $\tau_W = -i\text{Tr}\left(S^+ \frac{dS}{d\omega}\right) = \left(2|t|^2 \frac{d\varphi_t}{d\omega} + |r|^2 \frac{d\varphi_r}{d\omega} + |r'|^2 \frac{d\varphi_{r'}}{d\omega}\right) - i\frac{1}{2}\frac{d}{d\omega}(2|t|^2 + |r|^2 + |r'|^2)$ is no longer real. The real part of Wigner delay time is then $\text{Re}(\tau_W) = |t|^2 \frac{d\varphi_t}{d\omega} + |t'|^2 \frac{d\varphi_{t'}}{d\omega} + |r|^2 \frac{d\varphi_r}{d\omega} + |r'|^2 \frac{d\varphi_{r'}}{d\omega}$.

## APPENDIX D: DWELL TIME IN RANDOM ABSORBING SAMPLES

The dependence upon absorption strength of $\langle U \rangle$ and hence $\tau_D$ in 1D and in quasi-1D samples with $N = 8$ for different scattering strengths is shown in Fig. 17(a) and 17(b), respectively. There is a sharp drop in $\langle U \rangle$ for moderate absorption that increases with scattering strength and a gradual approach to the results in a uniform sample without scattering as absorption increases. In a uniform 1D sample, the energy excited from excitation from one side decays exponentially within the sample as $e^{-x/\ell_a}$ with $\ell_a = 2\gamma v_+$. Summing the integral of energy density for excitation from both sides gives, $U = \frac{1}{\gamma}\left(1 - e^{-2\gamma L/v_+}\right)$.

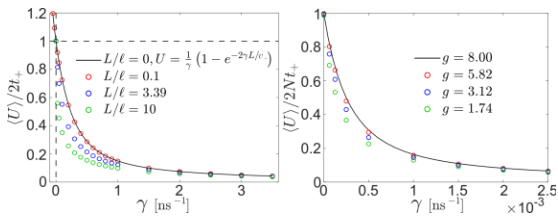

Fig 17. Energy excited in random medium vs. strength of absorption in samples with different scattering strengths. (a) Variation with absorption of a uniform sample and of random 1D samples of different scattering strength vs. loss and gain. (b) The same as (a) but for uniform and random quasi-1D samples with 8 channels for samples with different average transmittance, $\langle T \rangle = g$. For the uniform sample, the sample is perfectly transmitting so that $\langle T \rangle = N = 8$.

In lossless random media, the average reflection and transmission times are equal. But the reflection time falls as the loss increases, as can be seen in Fig. 3(d) for 1D media. Partial waves that penetrate more deeply into the sample and reside longer in the lossless sample are more strongly affected by absorption. As a result, the average reflection time falls, and $\langle U(x) \rangle$ falls towards the center of the sample. This is in accord with the greater suppression of the field due to modes with narrower linewidths, seen in Eq. (8a). Since the modal weights $|a_m(x)|^2$ for modes with narrower linewidths are further removed from the boundaries, the energy density inside the sample associated with such modes is more strongly suppressed and $\langle U(x) \rangle$ falls away from the boundaries.

When the absorption length falls below the transport mean free path, $\ell_a < \ell$, the wave behaves ballistically within an absorption length and $\langle U \rangle$ is dominated by the wave intensity in region in which scattering does not play a crucial role. Thus, $\langle U \rangle$ approaches the result for a sample without scattering. The approach occurs at smaller values of $\gamma$ in more weakly scattering samples with larger $\ell$ since the condition $\ell_a = \frac{v_+}{2\gamma} \sim \ell$ for a smaller value of $\gamma$.

The average transport properties of particles are the same as the average over a random ensemble of diffusing waves. The is no diffusive regime in 1D since the localization length is equal to the scattering length, $\xi = \ell$. Hence, the wave is either ballistic, $L < \ell$, or localized, $L > \xi$. A diffusive regime exists in quasi-1D media when $N > g > 1$. For $g \sim N$, transmission is nearly perfect, and the wave is ballistic. The results for $g = 3.12$ in Fig. 17(b) fall in the diffusive regime. Such results can be compared the results of a random walker such as insects that may become prey for larger insects or molecules that may dissociate.

## APPENDIX E: CALCULATION OF EXCITED ENERGY

The number of particles or total energy within a lossless medium with unit flux incident in all channels can be calculated in quantum systems following the Feshbach formalism [2]. The projection operators in the leads and scattering region, $P$ and $Q$ are $P = \sum_u \int dE |u_E\rangle\langle u_E|$, and $Q = \sum |m\rangle\langle m|$, where $|u_E\rangle$ and $|m\rangle$ indicate the continuous spectrum of channels in the leads and the discrete QNMs in the scattering region, respectively. We assume that the coupling between channels and modes is independent of frequency. Solving the quantum wave equation, $H\psi = E\psi$, the field excited within the sample is obtained by solving, $Q\psi = (E - H_{\text{eff}})^{-1}QHPu_E$, where $Q\psi$ is the projection of the wavefunction onto the scattering region. The operator $QHP$ describes the coupling between the scattering region and the leads. The effective Hamiltonian $H_{\text{eff}} = QHQ - \pi i \sum_u QHP|u_E\rangle\langle u_E|PHQ$ is the sum of the Hamiltonian for the scattering region and the coupling between the scattering region and its

surroundings.

The eigenvalues of $H_{\text{eff}}$ are the complex poles of the resonances of the open system, $H_{\text{eff}}|\varphi_{mr}\rangle = (E_m - i\gamma_m^0)|\varphi_{mr}\rangle$ and $\langle\varphi_{ml}|H_{\text{eff}} = \langle\varphi_{ml}|(E_m - i\gamma_m^0)$. Here, $|\varphi_{mr}\rangle$ and $\langle\varphi_{ml}|$ form the bi-orthogonal basis of the non-Hermitian Hamiltonian. The integral of the energy density within the sample is $U = 2\pi \sum_\psi <Q\psi|Q\psi>$. Plugging in the expression above for $Q\psi$ and making use of the relation for traces, $\text{Tr}(AB) = \text{Tr}(BA)$, we have $U = 2\pi\sum_\psi \langle Q\psi|Q\psi\rangle = 2\pi\sum_u\left\langle u_E\left|PHQ(E - H_{\text{eff}}^\dagger)^{-1}(E - H_{\text{eff}})^{-1}QHP\right|u_E\right\rangle = 2\pi\text{Tr}\left[(E - H_{\text{eff}}^\dagger)^{-1}(E - H_{\text{eff}})^{-1}\sum_u QHP|u_E\rangle\langle u_E|PHQ\right] = \text{Tr}\left[(E - H_{\text{eff}}^\dagger)^{-1}(E - H_{\text{eff}})^{-1}\left((H_{\text{eff}}^\dagger - H_{\text{eff}})/1i + 2\text{Im}(QHQ)\right)\right] = \sum_m[(E - E_m - i\gamma_m^0)^{-1}(E - E_m + i\gamma_m^0)^{-1}(2\gamma_m^0 + 0)] = \sum_m \frac{2\gamma_m^0}{(E-E_m)^2 + \gamma_m^{0\,2}} = \sum_m \frac{2\gamma_m^0}{(E-E_m)^2 + \gamma_m^{0\,2}}$. This gives Eq. (8a). However, this approach cannot be applied to find $U$ for classical waves experiencing loss or gain because the Hamiltonian $H$ is then not a Hermitian matrix and the relation $\sum_u QHP|u_E\rangle\langle u_E|PHQ = (H_{\text{eff}}^\dagger - H_{\text{eff}})/1i + 2\text{Im}(QHQ)$ is not valid.

## APPENDIX F: SPECTRA OF EXCITED FIELD AND COMPLETENESS OF QUASI-NORMAL MODES

The plots of $T$ and $U$ and $|E|^2$ in Figs. 4, 6 and 7 are based upon the fits to the field given by Eqs. (3a) and (8a) in samples with and without absorption, respectively. The poles used in fitting the spectra in Figs. 5-7 are shown in Tab. I in Appendix I. The blue and red curves in Fig. 18 are simulations of the real and imaginary part of the transmitted field at $x = 0.1L$ in samples with and without absorption whose square amplitude is shown in Fig. 6. The fit to the field of a unitary sample in Fig. 18(a) is obtained using Eq. 3(a) with the poles at $\omega_m^0 - i\gamma_m^0$ listed in Tab. I and the amplitudes $a_m^0(x)$ as fitting parameters. The fit shown as the dashed black curve in Fig. 18(a) is in excellent agreement with simulations. The dashed black curve in Fig. 18(b) for the absorbing sample is obtained using the amplitudes for the fit in Fig. 18(a) for the lossless sample in Eq. (8a).

In Fig. 19, we plot the amplitude squared of the field at $x = 0.01L$. Since $x$ is considerably smaller than the mean free path of $\ell = 0.29L$, the field is strongly correlated with the incident field, as was the case in Fig. 6. However, the quality of the fit to simulations shown as the black dashed curves in Fig. 29 is substantially degraded compared to the fit shown in Fig. 7 for $= 0.1L$. The fit improves considerably as indicated in the drop by more than two orders of magnitude in the value of chi-squared test $\frac{1}{N}\chi^2$ for the central part of the spectrum as the spectral window in which poles are found and used in the fit increases. This suggests that the modes are effectively complete, in accord with the fit of the square of the field amplitude at $x = 0.1L$ in Fig. 7, but that modes need to be determined in a larger spectral range. The need for a much larger spectral bandwidth to achieve a comparable fit of the field for points closer to the beginning of the sample may be associated with the increasing portion of the energy that has a short residence time in the sample.

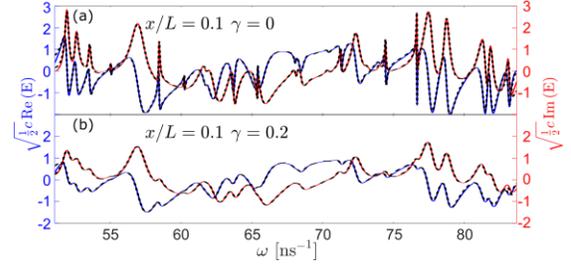

FIG. 18. Spectra of transmitted field. Real and imaginary parts of field for the sample of Figs. 4-7 for $x/L = 0.1$ calculated over a wide spectral range (a) without loss and (b) with loss of $\gamma = 0.2$.

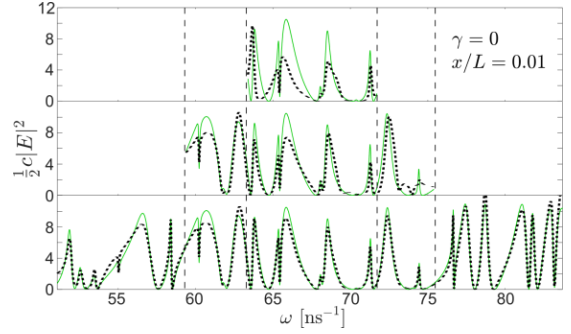

FIG. 19. Spectra of intensity near the input surface in different spectral windows. Simulations of $\frac{1}{2}c|E|^2$ (solid green curves) and the fit of the coherent sum in Eq. (8b) (black dashed curves) for spectra of increasing width for the same sample as in Figs. 6 and 7 for $x/L = 0.1$ without loss. The fits in the center of the spectra improve as the width of the spectral window increases but is of lower quality than fits for $x/L = 0.1$. The chi-squared test $\frac{1}{N}\chi^2 = \frac{1}{N}\sum_{i=1}^{N}\frac{(O_i - E_i)^2}{E_i}$ for the fit of $\frac{1}{2}c|E|^2$ in the central part with increasing width of the frequency window gives 37.9, 1.38 and 0.164.

## APPENDIX G: SOLUTION OF THE TIME-INDEPENDENT WAVE EQUATION WITH GAIN

## BEYOND THE LASING THRESHOLD

Time-independent transfer matrix simulations of spectra of $\tau_T, T$, and $U$ for a lossless 1D sample and for a sample with gain exceeding the leakage rate, $-\gamma = |\gamma| \geq \gamma_m^0$, for some modes are shown as the blue curves in Fig. 20(a) and 20(b), respectively. The expressions for all quantities in terms of the modal quantities found for the system below the modal lasing threshold are shown as the dashed black curves and seen to overlap the simulations even above the modal lasing threshold of $|\gamma| \geq \gamma_m^0$. However, these simulations give an unphysical reduction in energy in calculations and simulations based on the time-independent wave equation. The source of this result is the application of the time-independent wave equation to model wave propagation above the lasing threshold [99]. The solution of the wave equation in a sample with a uniform amplification rate above the modal lasing threshold is not stationary and the energy inside the sample would grow without limit. In addition, A proper account of wave propagation above threshold would require the solution of the time-dependent wave equation [99]. In addition, spontaneous emission, which produces an unbounded output even without an injected beam, and inevitable saturation of the gain medium, would need to be included.

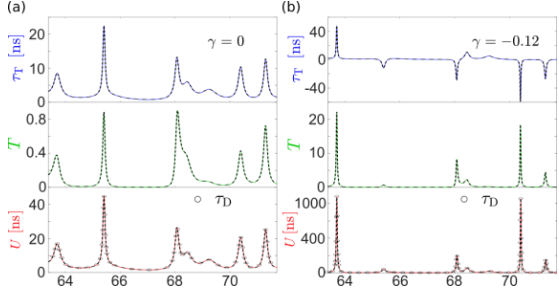

FIG. 20. Simulations of wave propagation in sample with and without gain beyond the lasing threshold. Simulations of transmission time, transmission, and total excited energy in the 1D sample of Fig. 5 for (a) a lossless sample and (b) a sample with $\gamma = -0.12 \text{ ns}^{-1}$. The spectra obtained for $U$ from Eqs. (5) and (6) and $\tau_D$ from Eq. (7) are identical.

As long as $\gamma_m > 0$, the scaling of the logarithm of transmission in random 1D media [95] is a simple generalization of the result of single parameter scaling in the lossless case [112]

$$\langle \ln T \rangle = -\frac{L}{\ell} - \frac{L}{\ell_a}, \quad (12a)$$

with $\frac{1}{\ell_a} = \frac{2\gamma}{v_+}$. The application of the time-independent wave equation to amplifying systems well above the lasing modal threshold, $-\gamma = |\gamma| \gg \gamma_m^0$, gives a dual symmetry between random ensembles with absorption and gain with the same values of $|\gamma|$ and the absorption length equals the gain length, $\ell_a = \ell_g = \frac{v_+}{2|\gamma|}$ [94,95,97], with

$$\lim_{L \to \infty} \frac{\partial \langle \ln T \rangle}{\partial L} = -\frac{1}{\ell} - \frac{1}{\ell_g}. \quad (12b)$$

To demonstrate the dual symmetry of Eqs. (12a) and (12b), we fix the sample length but change the absorption/gain rate. This give identical results for the variation of $\langle \ln T \rangle$ with $|\gamma|$ in the limit, $\gamma_m^0 + \gamma \to \gamma$. This gives

$$\lim_{|\gamma| \to \infty} \frac{\partial \langle \ln T \rangle}{\partial |\gamma|} = -2\pi\rho. \quad (12c)$$

With the same result for positive or negative values of $\gamma$. This can be derived as follows: $\lim_{|\gamma| \to \infty} \frac{\partial \langle \ln T \rangle}{\partial |\gamma|} = -\frac{\partial \left(\frac{L}{\ell_a}\right)}{\partial |\gamma|} = \frac{-\partial \left(2|\gamma|\frac{L}{v_+}\right)}{\partial |\gamma|} = -2\frac{L}{v_+} = -2t_+ = -2\pi\rho.$

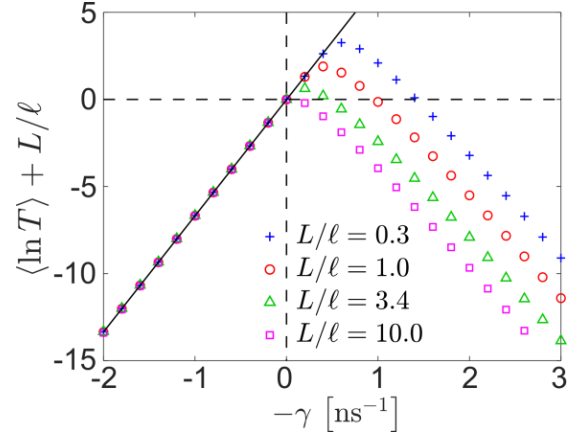

FIG. 21. Variation of average of logarithm of transmission with loss and gain. The straight line is $\langle \ln T \rangle + \frac{L}{\ell} = -\frac{L}{\ell_a} = \frac{L}{2v_+/\gamma} = -2t_+\gamma = -2\pi\rho\gamma.$

The results of simulation of the variation of $\langle \ln T \rangle + \frac{L}{\ell}$ with $-\gamma$ for different scattering strengths but fixed sample length are shown in Fig. 21. When only absorption is considered, $\langle \ln T \rangle + \frac{L}{\ell}$ varies with the absorption rate and the slope is independent of scattering strength. When gain is introduced into the system, the modal linewidth $\gamma_m = \gamma_m^0 + \gamma$ is positive below the modal lasing threshold. In this case, the transmission increases with gain and $\langle \ln T \rangle + \frac{L}{\ell} = -2\gamma\frac{L}{c}$, as in an absorbing medium. Beyond the localization threshold, a portion of the modes have negative linewidths so that $\langle \ln T \rangle + \frac{L}{\ell}$ does not increase as rapidly and finally decreases. Once the gain is strong enough that nearly all the modal linewidths are negative and $\gamma_m^0 + \gamma \to \gamma$, $\langle \ln T \rangle$ varies linearly with amplification rate $-\gamma$ and the slope approaches $-2\pi\rho$, in accord with the dual symmetry for strong absorption and gain. As the scattering strength increases, the turning point in the plot of $\langle \ln T \rangle + \frac{L}{\ell}$ occurs for smaller values of $-\gamma$, as seen in Fig. 21, as a result of the smaller values $\gamma_m^0$.

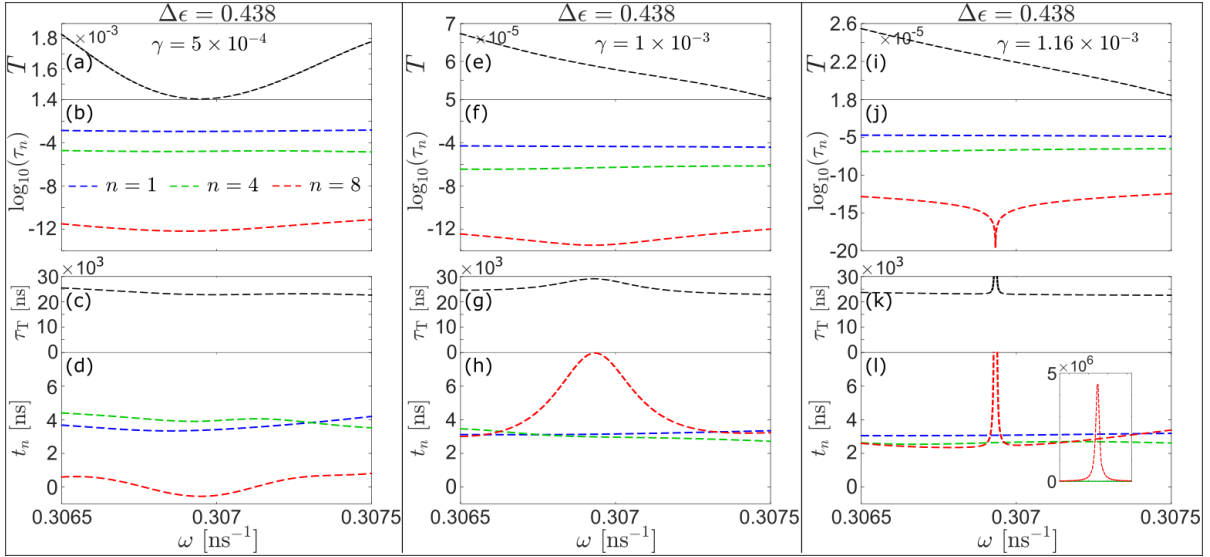

FIG. 22. Impact of transmission zeros on total and eigenchannel transmission and transmission time in strongly absorbing samples. Spectra are presented for the same configuration of real values of $\varepsilon$ presented in Fig. 13(e-h) with $\Delta\varepsilon = 0.438$ but with levels of absorption of (a-d) $\gamma = 5 \times 10^{-4}$, (e-h) $\gamma = 10^{-3}$, and (i-l) $\gamma = 1.16 \times 10^{-3}$, which are much higher than the levels of absorption in Fig. 1. In the first column, spectral features are broadened by absorption relative to spectra in Fig 14. Spectra are seen to sharpen in the second column with a clear dip in $\tau_8$ and a peak in $t_8$. Ultrasharp spectra are observed in the third column as the upper transmission zero of a conjugate pair is brought near the real axis from above.

## APPENDIX H: EIGENCHANNEL TRANSMISSION AND TRANSMISSION TIME IN STRONGLY ABSORBING SAMPLES

Here, we consider the impact of absorption upon spectra of transmission and transmission time for the sample with $\Delta\varepsilon = 0.438$ and the same spatial distribution of the real part of the dielectric function as studied in Figs. 14(e-h). The results for three different levels of absorption, all considerably higher than the level of absorption of $\gamma = 10^{-5}$ ns$^{-1}$, are shown in Fig. 22. The three dips seen in the spectrum of $t_8$ with $\gamma = 10^{-5}$ in Fig. 14(h) merge to form a single broad dip as the width of each of the zeros broadens by a factor of 50 in the sample with $\gamma = 5 \times 10^{-4}$ ns$^{-1}$, as seen in Fig. 22(d). At the same time, the background $t_8$ rises because the lowest transmission eigenchannel is no longer composed exclusively of far-off-resonance modes.

With a doubling the absorption rate of the field used in Figs. 22(a-d) to $\gamma = 10^{-5}$ ns$^{-1}$, a peak develops in $\tau_T$ and $t_8$, as seen in Figs. 22(g) and 22(h). The added absorption lowers both zeros of the conjugate pair. The peaks in $\tau_T$ and $t_8$ indicates that a zero in the upper half of the complex frequency plane has approached the real axis. The peak in the transmission time due to the transmission zero is not manifest in other transmission eigenchannels. The broad dip in $\tau_8$ seen in Fig. 22(f) arises since transmission is lowered in the eigenchannel with the lowest transmission eigenvalue as a transmission zero approaches the real axis. This transmission zero had not been manifest in spectra of transmission time and transmission at lower absorption levels because it was so far from the real axis.

When $\gamma = 1.16 \times 10^{-3}$ ns$^{-1}$, the transmission zero lies slightly above the real axis creating a sharp dip in $\tau_8$ (Fig. 22(j)) and a narrow peak in $\tau_T$ (Fig. 22(k)) and $t_8$ (Fig. 22(l)) in accord with Eq. (11). The full peak is seen in the inset in Fig. 22(l). The background in the spectrum of $t_N$ is not small relative to that for more highly transmitting eigenchannels, as is the case in weakly absorbing samples, because close lying modes contribute to the lowest transmission eigenchannel when absorption is strong.

## APPENDIX I: POLES AND ZEROS IN 1D AND QUASI-1D SAMPLE

Here we show all the poles and zeros used in Figs 5, 7, 9, 18-20 for 1D and Figs. 13(b), 14(a-d), 16 for quasi-1D in the complex frequency plan.

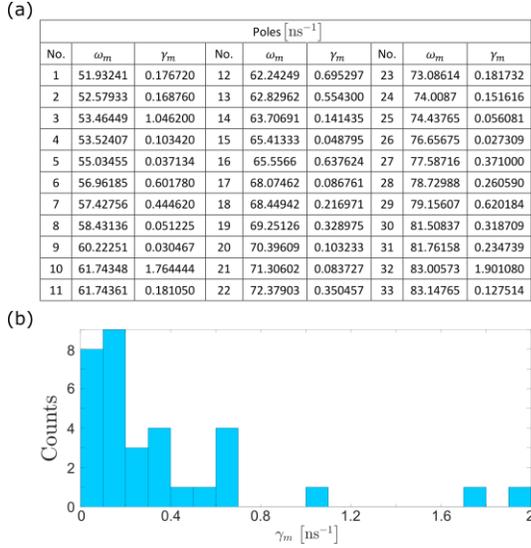

Tab. I. Poles in the complex frequency plane. (a) All the poles found for the unitary sample for the configuration for which simulations are shown in Figs. 5-7, 9, 18-20 in frequency range from [51.070, 83.698] ns$^{-1}$ and in half linewidths ranging from 0 to 2 ns$^{-1}$. (b) Histogram of the half-linewidths of the poles.

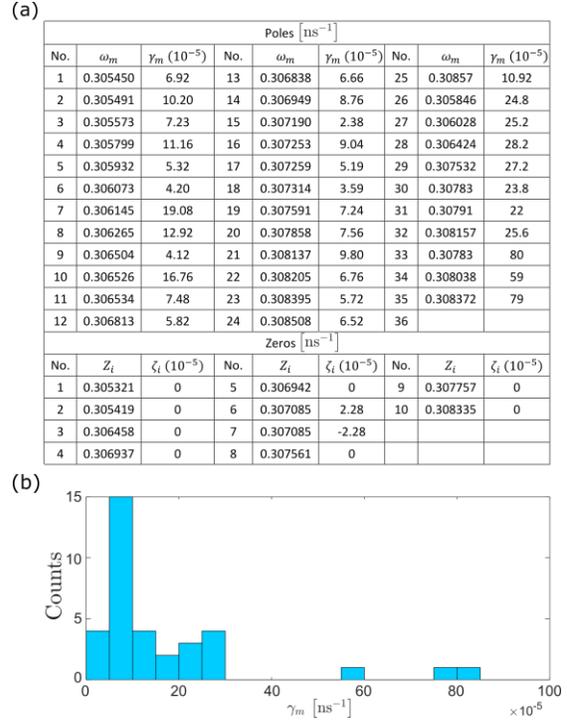

Tab. II. Poles and zeros in the complex frequency plane. (a) All the poles and zeros found from the phase map in the frequency range [0.3053, 0.3087] ns$^{-1}$ and in the imaginary part of the frequency $\zeta$ in the range $[-1 \times 10^{-3}, 1 \times 10^{-3}]$ ns$^{-1}$ in the sample configuration with $\Delta\varepsilon = 0.435$ for which data is shown in Figs. 13(b), 14(a-d), and 16. There are a total of 35 poles, 8 zeros on the real axis and 2 zeros off the real axis in this sector of the phase map. The poles from No. 1 to No. 25 are seen in the phase map shown in Fig.16. (b) Histogram of the half-linewidths of the poles.